\documentclass{article}%
\usepackage{amssymb}
\usepackage{amsmath}
\usepackage{revsymb}
\usepackage{eurosym}
\usepackage{graphicx}
\usepackage{amsfonts}
\usepackage[landscape,pdftex]{geometry}%
\providecommand{\U}[1]{\protect\rule{.1in}{.1in}}

\def\TextSymbolUnavailable#1{\textbf{???}}
\ifx\pdfoutput\relax\let\pdfoutput=\undefined\fi
\newcount\msipdfoutput
\ifx\pdfoutput\undefined\else
\ifcase\pdfoutput\else
\ifx\paperwidth\undefined\else
\ifdim\paperheight=0pt\relax\else\pdfpageheight\paperheight\fi
\ifdim\paperwidth=0pt\relax\else\pdfpagewidth\paperwidth\fi
\fi\fi\fi
\begin{document}

\title{Meromorphization\ of Large $N$ QFT}
\author{A.A.Migdal}
\maketitle

\begin{abstract}
We find general relations between RG equations and planar
unitarity-analyticity. These relations are summarized in meromorphization
procedure, generalizing the Pad\'{e} approximation in the limit of infinite order.

We also investigate confinement conditions for the mass spectrum in
asymptoticaly free QFT and lay down systematic framework for $\alpha$
expansion suggested in previous papers. The new relations for meromorphization
of symmetric conformal tensors are found, with resulting rich mass spectrum as
a function of spin $n.$

Explicit intergal representation for the triple string vertex $\Gamma$ is
found. This corresponds to resonanse theory with infinite number of masses and
Lagrangean
\[
\Phi Q\Phi+\Gamma\Phi^{3}%
\]

\end{abstract}
\tableofcontents
\listoffigures

\section{Introduction.}

\bigskip

Recently \cite{Migdal11}, we revived old approach to planar confining QFT
\cite{MigdalPade,MigdalAlpha}. The motivation for this revival was the
observation \cite{Low} that the $Ads_{5}/CFT$ theory being regularized by a
sharp cutoff in the vicinity of its $R_{4}$ boundary leads to the mass
spectrum exactly the same as suggested in \cite{MigdalPade,MigdalAlpha},
namely roots of Bessel functions. It was proven only for conserved currents
2-point functions of $CFT$, where there were no corrections to anomalous
dimensions of conformal operators. However, the regularization in
\cite{MigdalPade,MigdalAlpha} following solely from unitarity and analyticity,
there are strong reasons to believe that general formulas, with anomalous
dimensions as indexes of Bessel functions also apply to regularized $Ads/CFT$
theory. This would be an exciting phenomenon: exact relation which holds in
(regularized) $Ads/CFT$ theory to all orders in coupling constant. Recent
nonperturbative computations \cite{Kazakov} of anomalous dimensions in $N=4$
SYM theory as functions of coupling constant make this conjectured relation
even more exciting.

Can we now do something about the mass spectrum in confining $QFT$? At
phenomenological level we know that Pad\'{e} method works for QCD with vacuum
condensates (\cite{SVZ1,SVZ2}).

Before presenting the results of this paper, let me explain how and why the
mass spectrum is related to the OPE. It is relatively simple to go from the
spectrum to the OPE, but not so simple to go the other direction.

\bigskip

Take the sum of pole terms%
\begin{equation}
G(t)=\sum_{n=0}^{\infty}\frac{Z_{n}}{m_{n}^{2}-t}%
\end{equation}

and apply the Euler summation formula%
\begin{equation}
G(t)=\frac{1}{1-\exp(\partial_{n})}\left[  \frac{Z_{n}}{m_{n}^{2}-t}\right]
_{n=0}%
\end{equation}

This produces asymptotic expansion (with $B_{p}$ being Bernoulli numbers)
\begin{equation}
G(t)\rightarrow-\sum_{p=0}^{\infty}\frac{B_{p}}{p!}\left(  \partial
_{n}\right)  ^{p-1}\left[  \frac{Z_{n}}{m_{n}^{2}-t}\right]  _{n=0}%
\end{equation}

with the first term corresponding to replacement of discrete sum by an
integral, and the rest being corrections. We can introduce the density of the
spectrum, assuming we know analytic continuation of $Z_{n}$ and $m_{n}^{2}$ to
continuous values of $n$
\begin{equation}
\rho(m^{2})=\frac{dn}{dm^{2}},Z(m_{n}^{2})=Z_{n},
\end{equation}

and rewrite this expansion as an integral plus corrections%
\begin{equation}
G(t)\rightarrow\int_{m_{0}^{2}}^{\infty}ds\frac{\rho(s)Z(s)}{s-t}-\sum
_{p=1}^{\infty}\frac{B_{p}}{p!}\left(  \frac{1}{\rho(s)}\partial_{s}\right)
^{p-1}\left[  \frac{Z(s)}{s-t}\right]  _{s=m_{0}^{2}}%
\end{equation}

The first term would produce the continuum spectrum, corresponding to leading
order of asymptotic freedom.The higher Bernoulli terms will produce powerlike
corrections, proportional to negative integer powers of $t$. Those must
correspond to the higher terms of OPE, or else they must vanish. Take the
simplest possible Anzatz, the notorious psi function with%
\begin{equation}
Z_{n}=1,m_{n}^{2}=n+1
\end{equation}

It corresponds to constant spectral density, and so, the only corrections to
the leading term $\ln t$ will come out as integer negative powers. That
clearly contradicts the spectrum of OPE for large N YM, where there are no
conserved currents except Energy Tensor. All dimensions are known to get
perturbatively renormalized, so that there must be calculable fractional
powers of $\ln t$ in front of negative integer powers.

Such terms can only come from the integral term, due to some nontrivial
behavior of discontinuity%
\begin{equation}
\delta G(t)=\pi\rho(t)Z(t)
\end{equation}

which must posses all these powers of $t$ and $\ln t$. At the same time we
observe that all the Bernoulli terms must vanish identically, otherwise there
would be fake operators with integer dimensions in our theory. This provides
highly nontrivial restrictions on the mass spectrum $m_{n}$ and residues
$Z_{n}$.

Thus, we see that there are some hidden connections between the spectrum of
anomalous dimensions of perturbation theory and the physical mass spectrum,
and given the mass spectrum it is straightforward to compute the UV asymptotic
expansion and compare it with the OPE, and thus verify correctness of the
hypothesis about spectrum.

Our goal in this paper is to elaborate the same hidden connection between
masses and anomalous dimensions going the other way- from asymptotic freedom
to the physical spectrum. This way is much harder, but not impossible, as I
have shown 35 years ago.

\qquad

In this paper we continue investigation of Pad\'{e} regularization. We
rederive, debug and reinterpret old formulas of \cite{MigdalPade,MigdalAlpha},
and we make some new advances and insights. We are not going to assume
conformal symmetry but rather consider an arbitrary large $N$ QFT with
confinement, i.e. discrete spectrum of states in every channel. The specific
structure of $CFT$ and its $Ads$ correspondence will not be used here.

As it was discussed at length in \cite{MigdalPade,MigdalAlpha} one must use
the matrix Pad\'{e} approximant rather than scalar one. Nothing changes in
principle except necessity to keep track of order of matrix multiplication. We
carry out this matrix Pad\'{e} approximation to the end for the case of
arbitrary symmetric conformal tensors, where the 2-point function is the
matrix in space of $O(d)$ representation. As a result we get new rich mass
spectrum for such conformal tensors, overlooked in previous work.

The essence of our method is to impose on the n-point functions in momentum
space their correct analytic properties, i.e. meromorphicity. In case of
2-point function $G(t)$, where $t=-p^{2}$ we demand that it only has poles,
all located on the right semi-axis. We shall suppress the matrix indexes so
far and restore them later. We call this transformation
\textbf{meromorphization}:%
\begin{align}
M[G,Q,t]  &  =\frac{P(t)}{Q(t)},\label{Meromorphization}\\
P(t)  &  =G(t)Q(t)-\int_{0}^{\infty}\frac{ds}{\pi(s-t)}\delta_{s}G(s)Q(s).
\end{align}

where $\delta_{s}G(s)$ is discontinuity of $G(t)$ across the cut at positive
real axis $s>0$. The entire function $Q(t)$ here is chosen in such a way that
the dispersion integral here decreases faster than any power, i.e. all the
Laurent expansion coefficients of this integral at infinity must vanish:%
\begin{equation}
\int_{0}^{\infty}ds\delta_{s}G(s)Q(s)s^{n}=0,n=0,1,... \label{NoPowerTerms}%
\end{equation}

The motivation for this requirement is that these negative power terms in
asymptotic expansion of $G(t)$ come from the OPE, with coefficients
proportional to VEV of various operators with vacuum quantum numbers, such as
traces of powers of $F_{\mu\nu}$ in YM theory. We would like to preserve all
such terms while adding our corrections, so that these corrections are
exponentially small at large momenta. Note, that in case (which we would like
to eventually achieve) when the mass spectrum is given by the roots of $Q(t)$
the extra term we add will become identically zero, because $\delta
_{s}G(s)Q(s)\propto\delta(s-m^{2})Q(s)=0$ in this case.

In other words, \textbf{meromorphization is an identical transformation of the
full theory}, with the purpose to improve the perturbation expansion. The
physical meaning of meromorphization is to declare that the theory is an
infinite collection of free fields $\Phi_{i}(x)$ with various masses and
couplings to the gauge invariant YM sources $J_{k}(x)$ conjugate to operators
$O_{k}(x)$ in YM theory.%
\begin{equation}
S_{eff}=\int d^{4}x\sum_{ab}\Phi_{a}\widehat{M}_{ab}(i\nabla)\Phi_{b}+\Phi
_{a}\widehat{\Gamma}_{ab}(i\nabla)J_{b}%
\end{equation}

The mass spectrum is given by vanishing eigenvalues of infinite matrix
$\widehat{M}(p)$ and the 2-point function will factorize as $\widehat{\Gamma
}^{\dag}(p)\widehat{M}^{-1}(p)\widehat{\Gamma}(p)$. As it was discussed in the
previous paper, the Matrix Pad\'{e} approximation of $G$ provides precisely
the same representation, disguised in a form of right or left multiplication
\begin{equation}
G=\widehat{\Gamma}^{\dag}\widehat{M}^{-1}\widehat{\Gamma}=PQ^{-1}%
=\widetilde{Q}^{-1}\widetilde{P}.
\end{equation}
The asymptotic results at large order $M,N$ of the approximant are the same,
and they can be proven to be factorized as required by free field theory, with
real masses and real coupling constants. This remarkable coincidence follows
from the general theorems of Pad\'{e} theory for the Stiltjes matrix
functions. In particular, positivity and factorization was explicitly
demonstrated in the previous paper. There are more subtle theorems, such as
monotonous decrease of approximated mass spectrum with the rank $N$ of
Pad\'{e} approximant, and theorems about absolute convergence of approximant
in complex plane of $t$.

Do not let the word "approximation" mislead you. The so called approximant is
a rational matrix function, with coefficients, satisfying certain linear
matrix equations: so called Pad\'{e} equations. These equations are usually
solved numerically, which we are \textbf{not} going to do. We rather find
\textbf{exact} solutions of these equations for arbitrary order $M,N$ of
numerator and denominator polynomials. This solution represents certain
integral transformation of the matrix function $G(t)$ which transformation can
be defined exactly in every order of perturbation theory of OPE. Then we find
drastic simplification in the limit of large $M,N$ which allow us to go
further with the mass spectrum computation.

We start from Pad\'{e} approximation of $G(t)$ near large Euclidean point
$t=-\Lambda$ in the limit when the orders $M,N$ of Pad\'{e} approximant both
go to $\infty$ at fixed $R^{2}=\frac{2MN}{\Lambda}$. This $R$ has dimension of
length, so that conformal symmetry is explicitly broken. It may be restored in
the limit $R\rightarrow\infty$ . In case of precisely conformal field theory,
such as SYM dual to Ads string, we are looking for dynamic mechanisms
enhancing this conformal symmetry breaking (by VEV of composite fields) so
that it persists in a limit $R\rightarrow\infty$ . This is similar to
introduction of small magnetic field into ferromagnetic below Curie point to
obtain spontaneous magnetization in a limit of vanishing magnetic field.

In case of running coupling constant under consideration in this paper, the
conformal symmetry is broken already in the second loop. It would be
convenient, however, to separate the effects of renormalization of anomalous
dimensions, preserving conformal symmetry of YM theory, and effects from
running coupling constant leading to asymptotic freedom and - as we all \ hope
-- confinement. We study all these effects in some detail in this paper, to
review and extend the old results \cite{MigdalPade,MigdalAlpha}.

Few words about relation of two large space scales: confinement scale
$R_{QCD}=1/m_{0}$ which is exponentially large in perturbation theory, and our
regulator scale $R$ which we eventually must tend to infinity to recover
original theory. We start with the opposite limit of $R\ll R_{QCD}$ which is
like placing our QFT in a box size $R$ much smaller than confinement radius.
In this limit the spectrum is obviously discrete in every order of
perturbation expansion in running coupling $\lambda_{R}\sim1/\ln\left(
R_{QCD}/R\right)  $ . Then, we start increasing $R$ and develop the technique
of extrapolating the mass spectrum to the physical limit of $R\ \gg R_{QCD}$
in which limit we recover original theory.

The general theorems of Pad\'{e} theory for Stil'tjes functions are crucial
for this extrapolation. These theorems state that the position of poles of
approximant monotonously decrease as functions of its order. In our case this
means that mass spectrum is monotonously decreasing as $R$ passes the
confinement scale $R_{QCD}$ and goes to infinity. Nothing pathological like
Landau pole can happen with this definition of running scale. Confinement
would correspond to a finite limit of every mass in the spectrum at $R=\infty
$. We find the way to redefine the running coupling $\alpha$ so that it also
reaches the finite limit, namely $\alpha=1$. We argue that series expansion in
this running coupling $\alpha$ has finite radius of convergence, (precisely
$\alpha=1$). The expansion terms are calculable, so that the problem is to
extrapolate the series to the weak singularity (infinite first derivative) at
the convergence radius.

This approach was initiated in the old work (\cite{MigdalPade,MigdalAlpha})
but here we find new powerful method to compute terms of expansion, and also
correct some old errors.

We derive all the relevant meromorphization formulas from scratch in Appendix,
correcting some errors and typos along the way (including some which
propagated into the recent paper \cite{Migdal11}). The end results in the
limit $M,N\rightarrow\infty,R^{2}=\frac{2MN}{\Lambda}=const$ are quite simple.

First, let us study the relation between $P(t)$ and $Q(t)$. The two terms can
be combined as a single contour integral%
\begin{equation}
P(t)=\int_{C_{t}}\frac{dsG(s)Q(s)}{2\pi i(s-t)}.
\end{equation}

where $C_{t}$ encircles anti-clockwise both the pole at $s=t$ and
singularities of $G(s)$ at positive real axis.%

\begin{figure}[ptb]%
\centering
\ifcase\msipdfoutput
\includegraphics[
height=3.0571in,
width=3.9885in
]%
{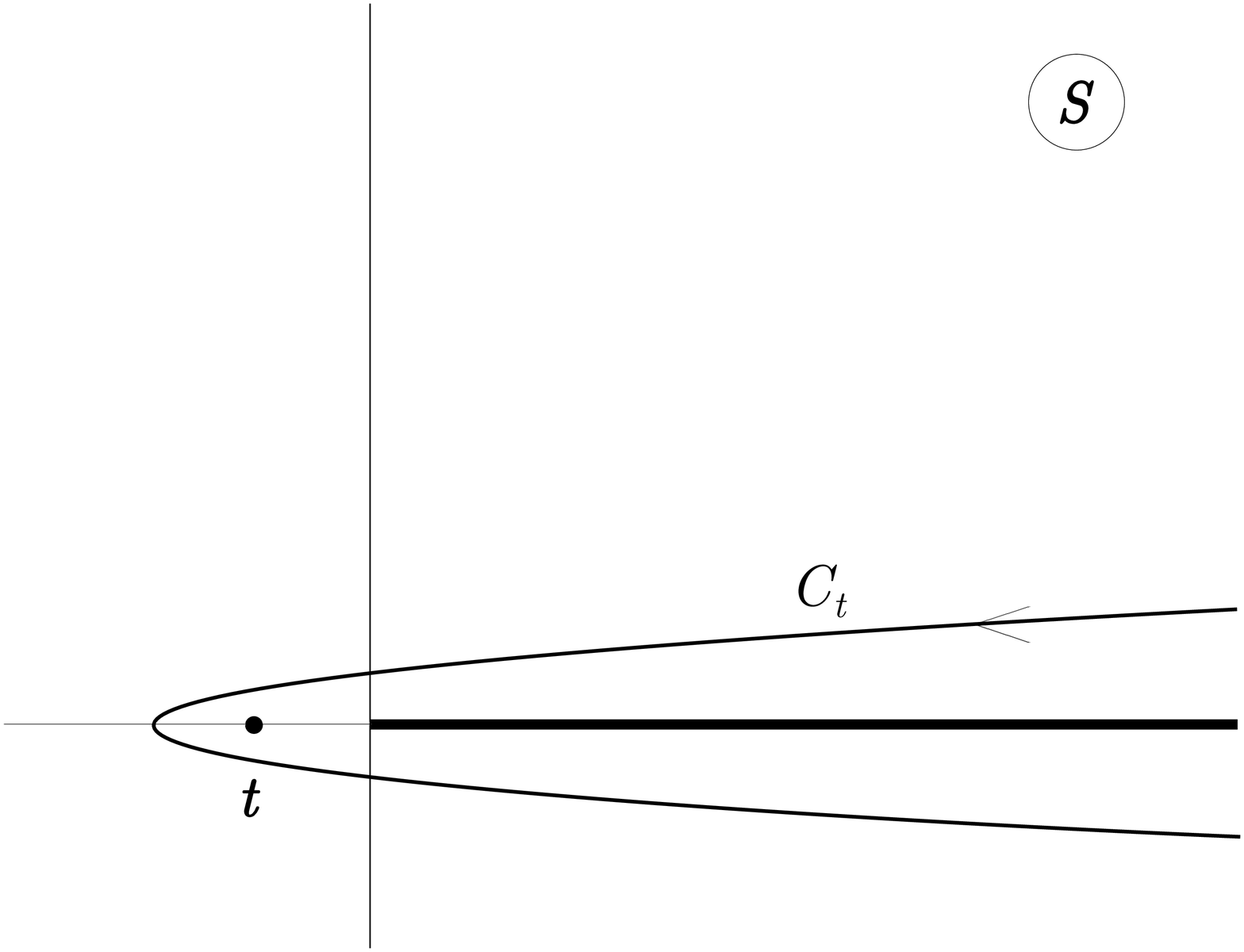}%
\else
\includegraphics[
height=3.0571in,
width=3.9885in
]%
{C:/whitepapers/Physics/Meromorphization/graphics/s-plane__1.pdf}%
\fi
\end{figure}

The contour cannot be closed in the left semiplane because of exponential
growth of $Q(s)$. The collision of singularities at $t\rightarrow0$ in this
contour integral does not lead to any singularities because they do not pinch
the contour. So, $P(t)$ is an entire function as well. Its expansion
coefficients $p_{n}$ in powers of $(-t)$ are given by the following contour
integral (in all subsequent formulas we choose $R$ as a unit of length)%
\begin{equation}
p_{n}=\int_{C_{0}}\frac{dsG(s)Q(s)}{2\pi i}(-s)^{-n-1}.
\end{equation}

Let assume that we know the Mellin transform of $G(s)$%
\begin{equation}
G(t)=\oint_{C_{F}}\frac{d\omega}{2\pi i}F(\omega)\left(  -t\right)  ^{\omega}.
\end{equation}

where $C_{F}$ encircles the singularities of $F(\omega)$. Substituting this
into relation for $p_{n}$ we get%
\begin{equation}
p_{n}=\oint_{C_{F}}\frac{d\omega}{2\pi i}F(\omega)\int_{C_{0}}\frac
{dsQ(s)}{2\pi i}(-s)^{\omega-n-1}%
\end{equation}

The last integral here is nothing but an expansion coefficient $q_{n}$ of
$Q(t)$ analytically continued to complex values of $n$%
\begin{align}
Q(t)  &  =\sum_{n=0}^{\infty}q_{n}\left(  -t\right)  ^{n},P(t)=\sum
_{n=0}^{\infty}p_{n}\left(  -t\right)  ^{n},\label{QDefs}\\
q_{\mu}  &  =\int_{C_{0}}\frac{dsQ(s)}{2\pi i}(-s)^{-\mu-1},\\
p_{\mu}  &  =\oint_{C_{F}}\frac{d\omega}{2\pi i}F(\omega)q_{\mu-\omega}.
\end{align}

This is quite a remarkable relation. Once the coefficients $q_{n}$ of the
entire function $Q(t)$ are known as analytic functions of index $n$ the other
entire function $P(t)$ is as good as known. We shall see below how this
relation works in practice, within planar graph expansion.

In general case, of many point function, it can be represented within planar
graph expansion as multiple Mellin integral \cite{NaturalLanguage} of various
planar kinematical invariants
\begin{align}
s_{ab}  &  =-\left(  \sum_{l=a}^{b}k_{l}\right)  ^{2},\\
G^{N}\left(  s_{..}\right)   &  =\int D\Omega(\omega)F^{N}\left(  \omega
_{..}\right)  \prod_{<ab>}\left(  -s_{ab}\right)  ^{\omega_{ab}}.
\label{MellinGN}%
\end{align}

Meromorphization in variable $s_{ab}$ would correspond to multiplication by
$Q(s_{ab})$ and contour integration as before. We will get multiple power
series expansion in numerator and product of $Q$ in denominator%
\begin{equation}
M\left[  G^{N},Q,s_{..}\right]  =\int D\Omega(\omega)F^{N}\left(  \omega
_{..}\right)  \prod_{<ab>}\frac{\sum_{n_{ab}=0}^{\infty}\left(  -s_{ab}%
\right)  ^{n_{ab}}q_{n_{ab}-\omega_{ab}}}{Q(s_{ab})}%
\end{equation}
Below, we carry out this procedure to the end for conformal 3-point vertex.

\section{Conformal Approximation.}

\bigskip

The equation for $q_{n}$ is less trivial than equation for $p_{n}$. Let us
assume that $F(\omega)$ is given by sum of two terms%
\begin{equation}
F(\omega)=\frac{r}{\omega-\gamma}+rS(\omega),\label{RSterms}%
\end{equation}
where $\gamma$ is the extreme right pole of $F(\omega)$ with the residue $r.$
In terms of $G(t)$ this corresponds to powerlike behavior. The second term
(which can also have singularities at $\omega=\gamma$) will be treated as
perturbation. This term, to be studied in the next Chapter, sums up all the
effects of running coupling constant, and starts from the second order in
perturbation expansion. As it follows from general relation derived in
Appendix , the coefficients $q_{\mu}$ satisfy the following equation
\begin{align}
q_{\mu} &  =\frac{1}{\Gamma(\mu+1)\Gamma(\mu+\gamma+1)}-\sum_{n=0}^{\infty
}S_{\mu n}q_{n},\label{qEquation}\\
p_{\mu} &  =rq_{\mu-\gamma}+r\oint_{C_{F}}\frac{d\omega}{2\pi i}%
S(\omega)q_{\mu-\omega}%
\end{align}

\begin{equation}
S_{\mu n}=\oint_{C_{S}}\frac{d\omega}{2\pi i}\digamma_{\gamma}\left(
\mu,n+\omega-\gamma\right)  S(\omega),
\end{equation}

\begin{equation}
\digamma_{\gamma}(a,b)=\frac{\sin(\pi b)\sin(\pi\left(  b+\gamma\right)
)}{\pi\sin(\pi\gamma)(b-a)}\frac{\Gamma(b+1)\Gamma(b+\gamma)}{\Gamma
(a+1)\Gamma(a+\gamma)} \label{DeltaSymbol}%
\end{equation}

The same formulas also provide analytic continuation for complex values of
$\mu$ .

Let us consider first the leading term, corresponding to conformal theory, and
let us now restore the length scale $R$%
\begin{align}
q_{\mu}^{0}  &  =\frac{1}{\Gamma(\mu+1)\Gamma(\mu+\gamma+1)},p_{\mu}^{0}%
=\frac{r}{\Gamma(\mu-\gamma+1)\Gamma(\mu+1)},\\
Q^{0}(t)  &  =(-t)^{-\frac{\gamma}{2}}I_{\gamma}\left(  2\sqrt{-tR^{2}%
}\right)  ,P^{0}(t)=r(-t)^{\frac{\gamma}{2}}I_{-\gamma}\left(  2\sqrt{-tR^{2}%
}\right)  .
\end{align}

The meromorphized 2-point function reduces to the ratio of Bessel functions%
\begin{equation}
G^{0}(t)=r\frac{(-t)^{\frac{\gamma}{2}}I_{-\gamma}\left(  2\sqrt{-tR^{2}%
}\right)  }{(-t)^{-\frac{\gamma}{2}}I_{\gamma}\left(  2\sqrt{-tR^{2}}\right)
}=r(-t)^{\gamma}\left(  1+\frac{2\sin(\pi\gamma)}{\pi}\frac{K_{\gamma}\left(
2\sqrt{-tR^{2}}\right)  }{I_{\gamma}\left(  2\sqrt{-tR^{2}}\right)  }\right)
.
\end{equation}

In special case of integer $\gamma\rightarrow n$ corresponding to conserved
currents, the residue $r$ must grow inversely proportional to $\sin(\pi
\gamma)$%
\[
r=\frac{r^{\prime}}{\sin(\pi\gamma)}%
\]

so that (up to irrelevant additive regular term $r(-t)^{n}$ ) the 2-point
function becomes%
\[
G^{0}(t)\rightarrow r^{\prime}\left(  \frac{1}{\pi}t^{n}\ln(-t)+\frac{2}{\pi
}(-t)^{n}\frac{K_{n}\left(  2\sqrt{-tR^{2}}\right)  }{I_{n}\left(
2\sqrt{-tR^{2}}\right)  }\right)  .
\]

with first term corresponding to its conformal limit.

The physical limit corresponds to large $R$ where the correction to power term
decays exponentially, as $\exp\left(  -4\sqrt{-tR^{2}}\right)  $, except at
positive $t$ where $G^{0}(t)$ has infinite number of poles with positive
residues. These properties follow from the general theorems of Pad\'{e} theory
for Stiltjes functions (for readers convenience the proof of positivity was
reproduced in \cite{Migdal11} ). The physical meaning is transparent: these
are the composite states of our large $N$ theory in conformal approximation.

Note that exponentially decaying corrections to the scaling limit in momentum
space correspond to analytic terms in coordinate space, going in powers of
$\frac{x}{R}$. Such corrections do not correspond to any physical operators in
OPE of $CFT$. From the point of view of confining QFT with conformal symmetry
explicitly broken by the beta function we do not see any problem with analytic
terms like these. In the limit of $R\rightarrow\infty$ these terms decay as
negative powers of $R$ , but after the summation of $\alpha$ perturbation
expansion (see below) these terms may stay finite.

\section{Perturbation Expansion for Mass Spectrum in Dimensional
Regularization.}

\bigskip

It is most convenient to study the running coupling constant with dimensional
regularization scheme. In that regularization, the 't Hooft coupling constant
$\lambda=Ng_{0}^{2}$ has dimension of $mass^{2\epsilon}$ where (in our
notations) $\epsilon=2-d/2$. The 2-point function expands in power series in
$\lambda$ with coefficients being some functions of $\epsilon$ times powers of
momentum as dictated by dimensional counting. Ignoring spins, we have%
\begin{equation}
G(t)=\sum_{k=1}^{\infty}\lambda^{k-1}f_{k}(\epsilon)t^{-k\epsilon}%
\end{equation}

The term $f_{k}$ represents the sum of $k-$loop planar diagrams for the
2-point function. This corresponds to Mellin transform $F(\omega)$ being a
simple sum of pole terms%
\begin{equation}
F(\omega)=\sum_{k=1}^{\infty}\frac{\lambda^{k-1}f_{k}(\epsilon)}%
{\omega+k\epsilon}%
\end{equation}

which correspond to $r=f_{1}(\epsilon),\gamma=-\epsilon$ in above equations.
The equation for denominator becomes%
\begin{align}
q_{\mu}  &  =\frac{1}{\Gamma(\mu+1)\Gamma(\mu-\epsilon+1)}-\sum_{n=0}^{\infty
}S_{\mu n}q_{n},\\
S_{\mu n}  &  =\sum_{m=1}^{\infty}\frac{f_{m+1}(\epsilon)}{f_{1}(\epsilon
)}\digamma_{-\epsilon}\left(  \mu,n-m\epsilon\right)  \left(  \lambda
R^{2\epsilon}\right)  ^{m}.
\end{align}

We can write down recurrent equations for expansion coefficients of $q_{\mu}$
in powers of the 't Hooft bare coupling.%
\begin{align}
q_{\mu}  &  =\sum_{m=0}^{\infty}q_{\mu}^{m}\left(  \lambda R^{2\epsilon
}\right)  ^{m},\\
q_{\mu}^{\left(  0\right)  }  &  =\frac{1}{\Gamma(\mu+1)\Gamma(\mu
-\epsilon+1)},\\
q_{\mu}^{\left(  r\right)  }  &  =-\sum_{m=1}^{r}\frac{f_{m+1}(\epsilon
)}{f_{1}(\epsilon)}\sum_{n=0}^{\infty}\digamma_{-\epsilon}\left(
\mu,n-m\epsilon\right)  q_{n}^{\left(  r-m\right)  }%
\end{align}

The expansion for the masses around the roots $m_{i}$ of Bessel function can
be done by iterating equation%
\begin{equation}
\frac{d\ln m_{i}}{d\lambda}\sum_{n=1}^{\infty}nq_{n}\left(  R^{2}m_{i}%
^{2}\right)  ^{n}=-\sum_{n=0}^{\infty}\frac{dq_{n}}{d\lambda}\left(
R^{2}m_{i}^{2}\right)  ^{n}%
\end{equation}

Note that (as it should have happened) the overall normalization of $G(t)$
corresponding to multiplicative renormalization constants of the operators,
dropped in our equation for the denominator. In particular, the one loop term
$f_{1}(\epsilon)$ had the pole at $\epsilon=0$ related to the logarithmic
divergence of the one loop integral in $d=4$ dimensions. Now this pole enters
denominator, so that, in effect, we have an extra factor of $\epsilon$ in the
numerator. This cancels the pole at $\epsilon=0$, coming from the kernel
$\digamma_{-\epsilon}\left(  \mu,n-m\epsilon\right)  $. Namely, there is the
factor $\sin(\pi\gamma)=-\sin(\pi\epsilon)$ in denominator. Now this pole at
$\epsilon=0$ is compensated by $f_{1}(\epsilon)$. So, the counting of
$\epsilon-$poles remains the same as in momentum space, though, of course
extra zeroth and positive powers of $\epsilon$ coming from the kernel, change
the resulting finite terms.

As we learned in the seventies from the famous 't Hooft's work, in order for
the observables to remain finite in the limit of $\epsilon=0$ all we need to
do is to renormalize the bare coupling%
\begin{equation}
\lambda=R^{-2\epsilon}\lambda_{R}\left(  1+c_{1}(\epsilon)\lambda_{R}%
+c_{2}(\epsilon)\lambda_{R}^{2}+...\right)
\end{equation}

The physical coupling $\lambda_{R}$ corresponds to space scale $R$ and is
supposed to remain finite at $\epsilon=0$. For that, the coefficients
$c_{k}(\epsilon)$ must have some poles, which are designed to cancel the
multiple poles in Laurent expansion of coefficient functions $f_{k}%
(\epsilon)\sim\epsilon^{-k}+...+\epsilon^{-1}+1$, after some other
multiplicative renormalization of $G(t)$. The existence of such universal
functions $c_{k}(\epsilon)$ that cancel poles in all 2-point functions is
called renormalizability. We take this for granted here. The UV
regularization, dimensional or otherwise should not be affected by the IR
regularization such as ours.

Moreover, we assume these functions known. Every multi-loop calculation of the
YM theory by necessity produces these functions, and also the similar
non-universal functions in the renormalization constants for the gauge
invariant operators. Naturally, we choose our parameter $R$ as a mass scale.
This will produce, after expansion of all the powers of $t$ and cancellation
of the poles in $\epsilon$ in every order in $\lambda_{R}$ some linear
combination of powers of $\ln R^{2}t$ in the momentum function $G(t).$

As for the expansion of our denominator $Q(t)$, it will produce some universal
numbers for expansion coefficients $q_{\mu}^{r}$ when re-expanded in running
coupling $\lambda_{R}$. The powers of $R$ will all cancel, as we have chosen
$R$ as our physical scale in the definition of the running coupling constant.
This follows now from trivial dimensional counting: there are no dimensional
parameters left.

Let us see how this expansion starts in the lowest order. We take the first
approximation, by multiplying $\digamma_{-\epsilon}\left(  \mu,n-\epsilon
\right)  $ by $q_{\mu}^{\left(  0\right)  }$. We get the following calculable sum

\begin{align}
q_{\mu}^{\left(  1\right)  }  &  =\frac{f_{2}(\epsilon)\sin(2\pi\epsilon)}{\pi
f_{1}(\epsilon)\Gamma(\mu+1)\Gamma(\mu-\epsilon)}\sum_{n=0}^{\infty}%
\frac{\Gamma(n-2\epsilon)}{\Gamma(n+1)}\frac{1}{(n-\epsilon-\mu)}\\
&  =\frac{f_{2}(\epsilon)}{f_{1}(\epsilon)}\frac{\sin\pi(\epsilon-\mu)}%
{\Gamma(\mu+1)\Gamma(\mu+\epsilon+1)\sin\pi(\epsilon+\mu)}%
\end{align}

At integer $\mu=m$%
\begin{equation}
q_{m}^{\left(  1\right)  }=\frac{f_{2}(\epsilon)}{f_{1}(\epsilon)}\frac
{1}{m!\Gamma(m+\epsilon+1)}\rightarrow\frac{f_{2}(\epsilon)}{f_{1}(\epsilon
)}\frac{1}{\left(  m!\right)  ^{2}}\left(  1-\epsilon\operatorname*{Psi}%
(m+1)\right)  .
\end{equation}

Comparing this with%
\begin{equation}
q_{m}^{\left(  0\right)  }=\frac{1}{m!\Gamma(m-\epsilon+1)}\rightarrow\frac
{1}{\left(  m!\right)  ^{2}}\left(  1+\epsilon\operatorname*{Psi}(m+1)\right)
.
\end{equation}

we see that the first term in $q_{m}^{\left(  1\right)  }$ leads to overall
renormalization, but the second term, being combined with the same term in
$q_{m}^{\left(  0\right)  }$ is equivalent to effect of anomalous dimension%
\begin{align}
q_{m}^{\left(  0\right)  }+\lambda_{R}q_{m}^{\left(  1\right)  }  &  =\frac
{Z}{m!\Gamma\left(  m+1+\gamma\right)  }+O(\lambda_{R}^{2}),\\
Z  &  =1+\frac{f_{2}(\epsilon)}{f_{1}(\epsilon)}\lambda_{R},\\
\gamma &  =-\epsilon+\frac{\epsilon f_{2}(\epsilon)}{f_{1}(\epsilon)}%
\lambda_{R}\rightarrow\frac{\epsilon f_{2}(\epsilon)}{f_{1}(\epsilon)}%
\lambda_{R}.
\end{align}

Therefore, the masses to the first order are given by the roots of $J_{\gamma
}(mR)$.

This is, of course, to be expected: the second loop adds one more logarithm
with $\frac{\epsilon f_{2}(\epsilon)}{f_{1}(\epsilon)}\lambda$ in front. This
is equivalent to anomalous dimension to this order. The running coupling
constant displays itself in the next loop. The nontrivial fact is that we
derived this from the Pad\'{e} equations, not from the conventional techniques
of summing up logarithms.

Higher terms are straightforward to generate, though I do not know whether
higher order sums would be analytically calculable, like in the first order.
These higher order terms involve some hypergeometric sums, which can be
expanded in powers of $\epsilon$ producing higher PolyGamma functions
$\operatorname*{Psi}(n+1,k)$ . The good news, however, is the convergence
factor we see in this perturbation expansion. At large $m$ our factors
decrease, at least they do so at any finite $\epsilon$.
\[
\digamma_{-\epsilon}\left(  \mu,n-m\epsilon\right)  \sim\frac{1}{m^{m\epsilon
}}%
\]

The problem of growth of planar graphs of high order (so called renormalons)
was not specific to four dimensions: at any finite $\epsilon$ 't Hooft
analytic argument would work as well. This argument does not apply to Pad\'{e}
regularization, precisely because it restores correct analytic properties of
2-point function. The 't Hooft's condensing singularities in complex $\lambda$
plane correspond to these same poles we have here, but old perturbation
expansion did not have. Nothing negative can be said about positions of these
poles from the point of view of analyticity: the renormalon argument loses its ground.

Another important comment. With the dimensional regularization scheme we
adopted here, there is no need to separate effects of multiplicative and
additive renormalizations. The 2-point function of composite fields has both
of these effects, as it is clear already in the leading order. There is a
logarithmic term, and logarithm transforms additively under rescaling of the
momentum cutoff. Therefore, the CS equations should be used with some care.
These equations do not apply to the 2-point function in momentum space, but
rather to coordinate space at non-coinsiding points. The ill-determined
subtraction polynomials in momentum, which reflect the additive
renormalization, drop after Fourier transformation. Another, more honest
alternative would be to write the CZ equations for the discontinuity of the
2-point function across the cut at positive $t$. This also eliminates all the
additive renormalizations. Discontinuity of the logarithm, for example, is
equal to $\pi$ and does not change with rescaling of the cutoff.

But in dimensional regularization there are no additive renormalizations to
begin with. Every Loop produces just a power of momentum with well defined
coefficient in front. This leads to great simplifications in meromorphization.
As we have seen above, these power terms are directly translated into
calculable terms in our Pad\'{e} equations. With powers of logarithms, one
would have to represent those as derivatives of powers to compute, which would
lead us back to dimensional regularization.

\section{Confinement and Beta Function.}

\bigskip

Let us now discuss the confinement condition, which arises in the limit
$R\rightarrow\infty$. According to the previous papers
\cite{MigdalAlpha,Migdal11} taking this limit of mass spectrum is equivalent
to its minimization over $R$ in view of monotonic decrease of masses as
functions of Pad\'{e} order. One can write each mass as follows (in convenient
logarithmic scale, in units of QCD mass squared scale $\Lambda_{QCD}$):%
\begin{equation}
\ln\left(  \frac{m_{i}^{2}}{\Lambda_{QCD}}\right)  =\lim_{R\rightarrow\infty
}\left(  -\ln\left(  R^{2}\Lambda_{QCD}\right)  +\rho_{i}(\lambda_{R})\right)
\end{equation}

where $\lambda_{R}$ is the running coupling, satisfying the RG flow equation
(negative sign because of space scale $R$ instead of mass scale)%
\begin{equation}
\frac{d\lambda_{R}}{d\ln\left(  R^{2}\right)  }=-\frac{1}{2}\beta(\lambda
_{R}), \label{RGFlow}%
\end{equation}

and $\rho_{i}(\lambda_{R})$ is given by our regularized perturbation
expansion, starting with constant:%
\begin{align}
\rho_{i}(\lambda_{R})  &  =\rho_{i}(0)+\lambda_{R}\rho_{i}^{\prime
}(0)+...;\label{RegPerturbation}\\
\rho_{i}(0)  &  =\ln r_{i}^{2}\\
J_{0}\left(  r_{i}\right)   &  =0;\\
\rho_{i}^{\prime}(0)  &  =-\frac{\epsilon f_{2}(\epsilon)}{f_{1}(\epsilon
)}\frac{\left.  \partial_{\gamma}J_{\gamma}(r_{i})\right\vert _{\gamma=0}%
}{r_{i}J_{0}^{\prime}(r_{i})}.
\end{align}

In order to compute these terms of expansion, one has to perturb the Bessel
solution for $Q$ by higher terms as explained in the previous Section. Also,
one should expand the anomalous dimensions, entering the indexes of the Bessel
functions, in power series in $\lambda_{R}$ . Combining all these
perturbations, we get the higher terms of expansion (\ref{RegPerturbation}).

Replacing the limit $R\rightarrow\infty$ by the extremum condition (valid for
our monotonous function!) we get%
\begin{equation}
\ln\left(  \frac{m_{i}^{2}}{\Lambda_{QCD}}\right)  =\min_{R}\left(  -\ln
(R^{2}\Lambda_{QCD})+\rho_{i}(\lambda_{R})\right)  \label{MassOriginal}%
\end{equation}

In order to make this limit non-singular we introduce Lagrange multiplier
$\alpha$ as follows\footnote{The transformation $\Phi(x)=\min_{y}(F(y)-xJ(y))$
is called Legendre transformation. It is well defined provided the extremum
equation $F^{\prime}(y)=xJ^{\prime}(y)$ has only one solution, which is our
case. Once this is true, there are several nice properties of Legendre
transform. In particular, as the value of the function at its extremum does
not depend upon the choice of the variable, the function $\Phi(x)$ is
invariant with respect to the change of initial variable $y:y\rightarrow
G(y)$. In particular, this means invariance with respect to renormalization
scheme. Another useful feature is the equation for derivative of $\Phi(x)$ :
$\Phi^{\prime}(x)=-J(y^{\ast})$, where $y^{\ast}(x)$ is the position of
extremum.The other terms in derivative $y^{\prime}(x)(F^{\prime}%
(y)-xJ^{\prime}(y))$ are identicaly zero due to the extremum condition. In our
case%
\[
x=\alpha^{2},J(y)=\ln(R^{2}\Lambda_{QCD}),
\]
and we can either use $\ln(R^{2}\Lambda_{QCD})$ as independent variable $y$,
in which case $J(y)=y$ and $F(y)=\rho(\lambda_{R})$, or else we can use
$y=\lambda_{R}$, in which case $F(y)=\rho(y)$ and%
\[
J(y)=2\int_{\lambda_{QCD}}^{\lambda_{R}}\frac{d\lambda}{\beta(\lambda)}.
\]
Results are independent of the choice of variable.}%

\begin{equation}
\ln\left(  \frac{m_{i}^{2}}{\Lambda_{QCD}}\right)  =\lim_{\alpha
\rightarrow1-0}\left\{  \min_{R}\left(  -\alpha^{2}\ln(R^{2}\Lambda
_{QCD})+\rho_{i}(\lambda_{R})\right)  \right\}  \label{MassAlpha}%
\end{equation}

Before we start working with this modified equation let us discuss in some
detail the crucial point of monotonous \ decrease of masses with respect to
$R$. For $\alpha=1$ it follows from the Pad\'{e} theorems. However, at
$\alpha<1$ this is no longer true. In case of confinement the function
$\rho_{i}(\lambda_{R})$ must grow at large $R$ precisely as $\ln(R^{2})$ to
cancel the first term. In other words, in case of confinement we have not only
the inequality%
\begin{equation}
\frac{d\rho_{i}(\lambda_{R})}{d\ln(R^{2})}\leq1
\end{equation}
which reflects monotonous decrease with $R$ of original mass $m_{i}^{2}$ in
(\ref{MassOriginal}), but we rather have an equality%
\begin{equation}
\left.  \frac{d\rho_{i}(\lambda_{R})}{d\ln(R^{2})}\right\vert _{R=\infty}=1
\end{equation}
This means, by continuity, that at least at large enough $R$
\begin{equation}
0<\frac{d\rho_{i}(\lambda_{R})}{d\ln(R^{2})}<1
\end{equation}
so that the minimum is unique. We cannot prove this positivity condition for
all $R$ but we can prove that it holds in the opposite asymptotically free
region of large negative $\ln(R^{2}\Lambda_{QCD})$. In this region the masses
are roots of Bessel functions $J_{\gamma}(mR)=0$, with index $\gamma$ being
anomalous dimension of the operator (up to $\lambda_{R}^{2}$ corrections). So,
we have%
\begin{equation}
\frac{d\rho_{i}(\lambda_{R})}{d\ln(R^{2})}=-\frac{1}{2}\beta(\lambda_{R}%
)\rho_{i}^{\prime}(\lambda_{R})\rightarrow-\frac{\lambda_{R}^{2}}{4}%
\beta^{\prime\prime}(0)\rho^{\prime}(0)=-\frac{\lambda_{R}^{2}}{2}%
\beta^{\prime\prime}(0)\gamma_{i}^{\prime}(0)\left.  \frac{d\ln r_{i}}%
{d\gamma}\right\vert _{\gamma=0}>0
\end{equation}
The derivatives of roots of Bessel function with respect to its index at zero
index are known to be positive, and so is $-\beta^{\prime\prime}(0)$. The
anomalous dimensions for some non-conserved currents are positive in the first
order, so that $\gamma^{\prime}(0)>0$ . We can choose which root we take as a
definition of our transformation from $R$ to $\alpha$. The parameter $R$ is
universal, so we can express it in terms of $\alpha$ from the lowest scalar
mass ($\gamma^{\prime}(0)>0$ in that case). Then we take ratios of
$m_{i}/m_{0}$ and re-expand the $\lambda_{R}$ expansion in series in $\alpha.$

As a result, we see that $\frac{d\rho_{0}(\lambda_{R})}{d\ln(R^{2})}$ is
positive in asymptotically free region as well as in the confinement region.
One could imagine some pathological behavior with this derivative changing
sign two times along the way from asymptotic freedom to the confinement
region. In that case, there may be several branches of the solution for
$\lambda_{R}$ as a function of $\alpha$. The best we can do is to compute the
perturbative branch, available to us, as a series in $\alpha$ and check
whether there are some phase transitions on the way from $\alpha=0$ to
$\alpha=1$ by estimating the radius of convergence.

Let us now dwell on the minimality condition, which can be rewritten in terms
of effective coupling $\lambda_{R}$

and the equation for the logarithm of mass which can be rewritten as follows%
\begin{equation}
\alpha^{2}=-\frac{1}{2}\beta(\lambda_{R})\rho_{i}^{\prime}(\lambda
_{R}),\label{MinimumEq}%
\end{equation}%
\begin{equation}
\ln\left(  \frac{m_{0}^{2}}{\Lambda_{QCD}}\right)  =\left[  -2\alpha^{2}%
\int_{\lambda_{QCD}}^{\lambda_{R}}\frac{d\lambda}{\beta(\lambda)}+\rho
_{0}(\lambda_{R})\right]  _{\alpha=1},\label{BetaEq}%
\end{equation}

In order to eliminate the unphysical variable $\lambda_{R}$ we must expand it
in power series in $\alpha$ by inverting (\ref{MinimumEq}) and substitute into
(\ref{BetaEq}). With the now popular choice of beta function, suggested in the
seventies (\cite{MigdalAlpha}, formula (1))%
\begin{align}
\Lambda_{QCD}  &  =\Lambda\exp\left(  -\frac{a}{\lambda_{0}}-b\ln\lambda
_{0}\right)  ,\label{MinimalBeta}\\
\frac{2}{\beta(\lambda)}  &  =-\frac{a}{\lambda^{2}}+\frac{b}{\lambda},\\
a  &  =\frac{96\pi^{2}}{11},b=\frac{102}{121}.
\end{align}

we have
\begin{equation}
\ln\left(  \frac{m_{0}^{2}}{\Lambda_{QCD}}\right)  =\left[  -\alpha^{2}\left(
\frac{a}{\lambda}+b\ln\lambda\right)  _{\lambda_{QCD}}^{\lambda_{R}}+\rho
_{0}(\lambda_{R})\right]  _{\alpha=1}.
\end{equation}

It is important to note that after taking minimum with respect to $R$, already
at arbitrary $\alpha<1$ the dimensional transmutation took place. The IR
cutoff $R$ disappeared, so that the mass has the correct QCD scale times some
function of dimensionless parameter $\alpha$%
\begin{equation}
m_{0}^{2}=\Lambda_{QCD}F(\alpha)
\end{equation}
These equations provide the basis for systematic expansion of $F(\alpha)$
using the perturbative expansions for $\beta(\lambda),\gamma(\lambda)$ and
$\rho_{0}(\lambda)$. The effective coupling starts linearly with $\alpha$%
\begin{align}
\lambda_{R}  & =\alpha\sqrt{\frac{a}{\rho_{0}^{\prime}(0)}}+O(\alpha^{2}),\\
F(\alpha)  & =r_{0}^{2}\left(  1+4\alpha\sqrt{a\rho_{0}^{\prime}(0)}\right)
+O(\alpha^{2})
\end{align}
After that, the dimensionless rations $m_{i}/m_{0}$ can be expanded in
$\alpha$, starting with ratios of roots of Bessel functions $J_{0}$%
\begin{align}
\ln\left(  \frac{m_{i}}{m_{0}}\right)   &  =\frac{1}{2}\left(  \rho
_{i}(\lambda_{R})-\rho_{0}(\lambda_{R})\right)  \\
&  =\ln\left(  \frac{r_{n}}{r_{0}}\right)  +\frac{1}{2}\alpha\sqrt{\frac
{a}{\rho_{0}^{\prime}(0)}}\left(  \rho_{i}^{\prime}(0)-\rho_{0}^{\prime
}(0)\right)  +O(\alpha^{2}),\\
J_{0}(r_{i}) &  =0;\\
\rho_{i}^{\prime}(0) &  =\gamma_{i}^{\prime}(0)\left.  \frac{d\ln r_{i}%
}{d\gamma}\right\vert _{\gamma=0}%
\end{align}
Since $\alpha$ starts linearly in $\lambda$ we may view this $\alpha$ as a
physical running coupling constant. The advantage of this redefinition of
coupling constant is obvious: unlike original coupling $\lambda$ this $\alpha$
tends to $1$ in the strong coupling limit $R=\infty$. In the weak coupling
limit $\alpha$ goes to zero as $-1/\ln R$ , same as $\lambda_{R}$.

The reader may wonder: how did we get around the notorious Landau pole? The
effective coupling $\lambda_{R}$ can have any singularities as a function of
$R$, it can have a pole, or even a branchpoint such that one would not be able
to continue $\lambda_{R}$ beyond some value of $R$ without getting imaginary
part\footnote{This is what happens with the popular minimal choice of beta
function at $\lambda crit=\frac{176}{17}\pi^{2}$.}.These troubles do not
reflect the physics of our system but rather the poor choice of effective
coupling. We \textbf{know} from Pad\'{e} theorems that $\ln(mass)$ decrease
monotonously as functions of $\ln R$ so we can perform Legendre transformation
and rely on nice properties of masses as functions of Legendre parameter
$\alpha^{2}$.

In fact we can plot this dependence qualitatively, as it was done in my old
papers. This follows from the famous Legendre formula, in our case%
\begin{equation}
\frac{d\ln(m_{0}^{2})}{d\alpha}=-2\alpha\ln(R^{2}\Lambda_{QCD})=-4\alpha
\int_{\lambda_{R}}^{\lambda_{QCD}}\frac{d\lambda}{\beta(\lambda)}%
\end{equation}

At small $\alpha$ the integral is dominated by the lower end, and produces
$\frac{-a}{\lambda_{R}}\propto\frac{-1}{\alpha}$ which cancels $\alpha$ and
leads to a positive finite limit.%
\begin{equation}
\frac{d\ln m_{0}^{2}}{d\alpha}\rightarrow\frac{4\alpha a}{\lambda_{R}}%
=4\sqrt{a\rho_{0}^{\prime}(0)}>0;\alpha\rightarrow0.
\end{equation}

It then reaches the maximum at $\Lambda_{QCD}$%
\begin{equation}
\frac{d\ln m_{0}^{2}}{d\alpha}=0,R^{2}=\frac{1}{\Lambda_{QCD}};
\end{equation}

This is all what happens with our theory at would-be Landau pole. This is a
position of the maximum of each mass in the spectrum as function of our
physical coupling $\alpha$. After this the mass goes down and reaches finite
limit at $\alpha=1$. However, this is a singularity, as the derivative%
\[
\frac{d\ln m_{0}^{2}}{d\alpha}=-\infty;\alpha=1
\]

\bigskip

Here is the simplest function with such a behavior:%

\begin{figure}[ptb]%
\centering
\ifcase\msipdfoutput
\includegraphics[
height=2.3134in,
width=3.659in
]%
{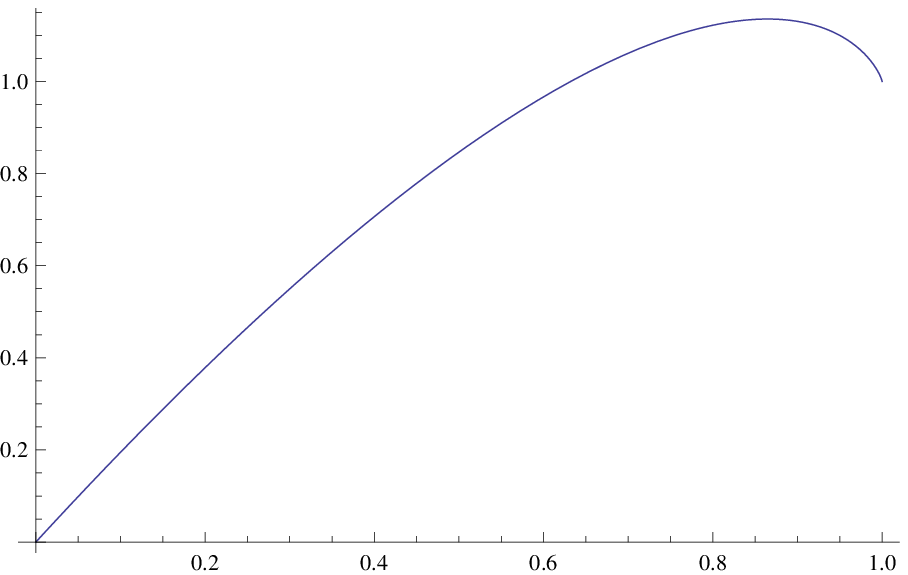}%
\else
\includegraphics[
height=2.3134in,
width=3.659in
]%
{C:/whitepapers/Physics/Meromorphization/graphics/mass__2.pdf}%
\fi
\caption{$\ln m^{2}=%
\protect\begin{array}
[c]{ccc}%
\alpha-(1-\alpha)\ln(1-\alpha) & ; & 0<\alpha<1
\protect\end{array}
$}%
\end{figure}

The extrapolation to a singular value $\alpha=1$ can be done by the continuos
fraction made from available few expansion terms (hopefully, $10$ at modern
level of analytic computations of the Feynman graphs of massless theory). The
convergence of continuos fraction at $\alpha=1$ to exact value $m=1$ for this
prototype function goes as follows (for various orders of continuous fraction)%

\[%
\begin{tabular}
[c]{ll}%
$order$ & $\ln m(\alpha=1)$\\
$2$ & $2$\\
$3$ & $1.6$\\
$4$ & $1.25$\\
$5$ & $1.17391$\\
$6$ & $1.11111$\\
$7$ & $1.08485$\\
$8$ & $1.0625$\\
$9$ & $1.05049$\\
$\infty$ & $1.0$%
\end{tabular}
\
\]

Let us also discuss the case of perturbations around the nontrivial fixed
point $\lambda^{\ast}$. In this case the beta function vanishes linearly so
one has to use another parameter $\xi=\sqrt{\alpha}$.%
\begin{equation}
\frac{1}{2}\beta(\lambda)\partial_{\lambda}t_{0}+\xi t_{0}=0,\beta
(\lambda)=(\lambda-\lambda^{\ast})\beta^{\prime}+O\left(  (\lambda
-\lambda^{\ast})^{2}\right)
\end{equation}

and expand in series of $\xi$. The rest of the arguments goes the same way,
with replacement of $\alpha^{2}\rightarrow\xi$. So, the $CFT$ perturbed around
the fixed point by a running coupling constant can confine as well, and the
above expansion can be used to compute its spectrum in terms of expansion in
powers of $\xi$.

\bigskip

\section{Meromorphization of Conformal Vertex}

\bigskip

From the point of view of string theory (or, better to say, dual resonance
theory, as we do not know nor we need to know explicit string model), the
denominator $Q$ provides quadratic part of the effective Lagrangian%
\begin{equation}
\sum_{ab}\Phi_{a}Q_{ab}\left(  \nabla^{2}\right)  \Phi_{b}.
\end{equation}

We ignore here the tensor structure, which in momentum space depend of
$n_{\mu}=\frac{p_{\mu}}{\left\vert p\right\vert }$. In order to achieve
analyticity, corresponding power of $p^{2}$ must be present in $Q$ to cancel
kinematical singularities of $\left\vert p\right\vert =\sqrt{-t}$ in
denominator of $n_{\mu}$ . We shall not go into these details in present paper
for clarity of presentation. We shall take here normalization of 2-point
function as pure power of $x$ in coordinate space, so that in momentum space
(for scalar case)%
\begin{equation}
G(-k^{2})=\int d^{d}xx^{-2{}\Delta}\exp(ikx)=\sigma(\nu)k^{2\nu};\sigma
(\nu)=\frac{\pi^{\frac{d}{2}}2^{-d-2\nu}\Gamma\left(  -\nu-\frac{d}{2}\right)
}{\Gamma(d+\nu)},\nu=\Delta-\frac{d}{2}.
\end{equation}

Let us now introduce some cubic interaction%

\begin{equation}
\sum_{abc}\Gamma_{abc}\left(  -i\partial_{1},-i\partial_{2},-i\partial
_{3}\right)  \Phi_{a}(1)\Phi_{b}(2)\Phi_{c}(3)
\end{equation}

and compare the conformal 3-point function%

\begin{equation}
\left\langle O_{i}(x_{1})O_{j}(x_{2})O_{k}(x_{3})\right\rangle =C_{ijk}%
x_{12}^{-\Delta_{i}-\Delta_{j}+\Delta_{k}}x_{23}^{-\Delta_{j}-\Delta
_{k}+\Delta_{i}}x_{31}^{-\Delta_{k}-\Delta_{i}+\Delta_{j}}. \label{CFTVertex}%
\end{equation}
to the tree diagram of this $\Phi^{3}$ theory. We need to go to momentum
space, meromorphize and multiply by three propagators $Q$ to obtain the vertex
$\Gamma_{abc}(q_{1},q_{1},q_{3})$ as a function of external momenta $q_{i}$.
Ignoring tensor structures and skipping indexes $a,b,c$ we have triangle
diagram with power propagators (here and below we denote by $i,j,k$ the cyclic
ordered indexes $1,2,3$)%

\begin{equation}
G(s_{1},s_{2},s_{3})=C_{123}\sigma\left(  \frac{\alpha_{1}}{2}\right)
\sigma\left(  \frac{\alpha_{2}}{2}\right)  \sigma\left(  \frac{\alpha_{3}}%
{2}\right)  \int\frac{d^{d}q}{\left(  2\pi\right)  ^{d}}\left\vert
q-q_{1}\right\vert ^{\alpha_{1}}\left\vert q-q_{2}\right\vert ^{\alpha_{2}%
}\left\vert q-q_{3}\right\vert ^{\alpha_{3}};s_{i}=q_{i}^{2};\alpha_{i}%
=\Delta_{j}+\Delta_{k}-\Delta_{i}-d.
\end{equation}

Exponentiating these power propagators we get%
\begin{equation}
G(s_{1},s_{2},s_{3})=C_{123}\left(  \prod_{i=1}^{3}\frac{\sigma\left(
\frac{\alpha_{i}}{2}\right)  }{\Gamma\left(  -\frac{\alpha_{i}}{2}\right)
}\int_{0}^{\infty}\frac{dx_{i}}{x_{i}^{1+\alpha_{i}/2}}\right)  \int%
\frac{d^{d}q}{\left(  2\pi\right)  ^{d}}\exp\left(  -\sum_{i=1}^{3}%
x_{i}\left(  q-q_{i}\right)  ^{2}\right)  ,
\end{equation}

The Gaussian integral over $k$ is straightforward. After some algebra (thanks
to Mathematica$^{\texttrademark}$) we get simple expression%
\begin{equation}
G(s_{1},s_{2},s_{3})=\frac{C_{123}}{\left(  2\sqrt{\pi}\right)  ^{d}}\left(
\prod_{i=1}^{3}\frac{\sigma\left(  \frac{\alpha_{i}}{2}\right)  }%
{\Gamma\left(  -\frac{\alpha_{i}}{2}\right)  }\int_{0}^{\infty}\frac{dx_{i}%
}{x_{i}^{1+\alpha_{i}/2}}\right)  \left(  \sum_{i=1}^{3}x_{i}\right)
^{-\frac{d}{2}}\exp\left(  -\sum_{i=1}^{3}s_{i}x_{j}x_{k}\right)  ,
\end{equation}

This integral further simplifies by the change of variables from $x_{i}$ to
\begin{equation}
y_{i}=x_{j}x_{k};\frac{D(y_{1},y_{2,}y_{3})}{D(x_{1},x_{2,}x_{3})}=2x_{1}%
x_{2}x_{3}=2\sqrt{y_{1}y_{2}y_{3}};x_{i}=\frac{\sqrt{y_{1}y_{2}y_{3}}}{y_{i}}.
\end{equation}

\begin{align}
\mu_{i}  &  =\frac{\Delta_{1}+\Delta_{2}+\Delta_{3}}{4}-\Delta_{i}%
,\label{MuDef}\\
\left(  \prod_{i=1}^{3}\int_{0}^{\infty}\frac{dx_{i}}{x_{i}^{1+\alpha_{i}/2}%
}\right)  \left(  \sum_{i=1}^{3}x_{i}\right)  ^{-\frac{d}{2}}\exp\left(
-\sum_{i=1}^{3}s_{i}x_{j}x_{k}\right)   &  =\frac{1}{2}\left(  \prod_{i=1}%
^{3}\int_{0}^{\infty}dy_{i}y_{i}^{\mu_{i}-1}\exp\left(  -s_{i}y_{i}\right)
\right)  \left(  \sum_{i=1}^{3}\frac{1}{y_{i}}\right)  ^{-\frac{d}{2}},
\end{align}
\qquad

Let us now use Mellin integral, summing up multinomial expansion%
\begin{equation}
\left(  \sum_{i=1}^{3}\frac{1}{y_{i}}\right)  ^{-\frac{d}{2}}=\frac{1}%
{\Gamma\left(  \frac{d}{2}\right)  }\oint_{C_{\Gamma}}D\Omega\prod_{i=1}%
^{3}\Gamma(\omega_{i})y_{i}^{\omega_{i}};D\Omega=\frac{d\omega_{1}d\omega_{2}%
}{\left(  2\pi i\right)  ^{2}};\omega_{3}=\frac{d}{2}-\omega_{1}-\omega_{2}.
\end{equation}

with integration contours $C_{\Gamma}$ encircling poles of $\Gamma(\omega)$ at
$\omega=0,-1,...$. Integral is, in fact, symmetric with respect to the choice
of two integration variables out of $\omega_{1},\omega_{2},\omega_{3}$.
\begin{align}
G(s_{1},s_{2},s_{3})  &  =g_{123}\oint_{C_{\Gamma}}D\Omega\prod_{i=1}%
^{3}\Gamma(\omega_{i})\int_{0}^{\infty}dy_{i}y_{i}^{\omega_{i}+\mu_{i}%
-1}e^{-s_{i}y_{i}}=g_{123}\oint_{C_{\Gamma}}D\Omega\prod_{i=1}^{3}%
s_{i}^{-\omega_{i}-\mu_{i}}\Gamma(\omega_{i})\Gamma\left(  \mu_{i}+\omega
_{i}\right)  ,\\
g_{123}  &  =\frac{C_{123}}{2\Gamma\left(  \frac{d}{2}\right)  \left(
2\sqrt{\pi}\right)  ^{d}}\prod_{i=1}^{3}\frac{\sigma\left(  \frac{\alpha_{i}%
}{2}\right)  }{\Gamma\left(  -\frac{\alpha_{i}}{2}\right)  }.
\end{align}

This expression defines analytic continuation to complex values of $s_{i}$.
The remaining steps follow meromorphization of the 2-point function,
independently for each variable $s_{i}$ . We have for the corresponding Taylor
coefficient in terms of $s_{i}$ :
\[
\int_{C_{0}}\frac{ds_{i}Q_{i}(-s_{i})}{2\pi i}(s_{i})^{-n_{i}-\omega_{i}%
-\mu_{i}-1}=q_{i,n_{i}+\mu_{i}+\omega_{i}}%
\]

and finally, for the meromorphized triple vertex (with $\mu_{i}$ defined in
(\ref{MuDef}) ):%
\begin{align}
\Gamma(s_{1},s_{2},s_{3})  &  =M[G,Q,s_{1},s_{2},s_{3}]Q(s_{1})Q(s_{2}%
)Q(s_{3})=\sum_{n_{1},n_{2},n_{3}\geq0}\gamma_{n_{1},n_{2},n_{3}}s_{1}^{n_{1}%
}s_{2}^{n_{2}}s_{3}^{n_{3}},\\
\gamma_{n_{1},n_{2},n_{3}}  &  =g_{123}\oint_{C_{\Gamma}}\frac{d\omega_{1}%
}{\left(  2\pi i\right)  }\oint_{C_{\Gamma}}\frac{d\omega_{2}}{\left(  2\pi
i\right)  }\prod_{i=1}^{3}q_{i,n_{i}+\mu_{i}+\omega_{i}}\Gamma(\omega
_{i})\Gamma\left(  \mu_{i}+\omega_{i}\right)  ;\omega_{3}=\frac{d}{2}%
-\omega_{1}-\omega_{2},
\end{align}

This is our goal: triple vertex defined as an entire function of each
variable. The matrix indexes are to be inserted in obvious places, we omit
them for brevity. The $q$ coefficients inside the integral decrease as square
of Gamma function of its arguments, providing decrease of the integral at
large $n_{i}$. This double contour integral can be reduced to double expansion
over residues at $\omega_{1}=-l_{1,}\omega_{2}=-l_{2}$ of gamma functions
$\Gamma(\omega_{1})\Gamma(\omega_{2})$ in left semiplanes%
\begin{equation}
\gamma_{n_{1},n_{2},n_{3}}=g_{123}\sum_{l_{1},l_{2}=0}^{\infty}\frac
{(-1)^{l_{1}+l_{2}}}{l_{1}!l_{2}!}\Gamma(\omega_{3})\prod_{i=1}^{3}%
q_{i,n_{i}+\mu_{i}+\omega_{i}}\Gamma\left(  \mu_{i}+\omega_{i}\right)
,\omega_{1}=-l_{1},\omega_{2}=-l_{2},\omega_{3}=\frac{d}{2}+l_{1}+l_{2}.
\label{GammaSum}%
\end{equation}
In particular, in conformal limit for $q$ we obtain infinite sum of ratios of
Gamma functions:%
\begin{equation}
\gamma_{n_{1},n_{2},n_{3}}^{0}=g_{123}\sum_{l_{1},l_{2}=0}^{\infty}%
\frac{(-1)^{l_{1}+l_{2}}}{l_{1}!l_{2}!}\Gamma(\omega_{3})\prod_{i=1}^{3}%
\frac{\Gamma\left(  \mu_{i}+\omega_{i}\right)  }{\Gamma(1+n_{i}+\mu_{i}%
+\omega_{i})\Gamma\left(  1+n_{i}+\gamma_{i}(\lambda)+\mu_{i}+\omega
_{i}\right)  } \label{GammaIntegral}%
\end{equation}

In general case of running coupling constant, the proper way of
meromorphization of the vertex function in perturbation expansion is to
represent Feynman graphs as Mellin transforms \cite{NaturalLanguage} and
meromorphize the powers of external variables $s_{i}$ as we did \ above, using
$q_{\mu}$.

\bigskip

\section{Spins and Operator Mixing}

\bigskip

So far we ignored the operator mixing, and studies idealized version of planar
QFT, without spins. Let us now take both of these effects into consideration.
As it was suggested in old papers \cite{MigdalAlpha} and reiterated recently
\cite{Migdal11} the proper framework, automatically preserving planar
unitarity-analyticity is given by large order limit of matrix Pad\'{e}
approximant. We are going to refer to this limit as \textbf{Matrix
Meromorphization}. Nothing changes in the general formulas of the
Introduction, as long as we treat $G,P,Q$ as infinite matrices in Hilbert
space of composite fields made of quarks and gluons. It will become the
Hilbert space of free composite particles of our planar QFT.

Let us now go into details of this Matrix Meromorphization. The matrix
$\mathbf{G}$ is acting in space of irreducible tensors of space-time symmetry
group $O(d)$ (we are working in Euclidean space so far, but we do not assume
conformal symmetry). There are tensor indexes for each of two operators
averaged in $\mathbf{G}=\left\langle O_{1}O_{2}\right\rangle $ . The tensor
$\mathbf{G}$ is made of products of $\varepsilon_{\mu\nu\lambda\rho...}$,
$\delta_{\mu\nu}$ and $\widehat{k}_{\mu}=\frac{k_{\mu}}{|k|}$ with scalar
functions of $t=-k^{2}$ in front of these invariant tensors. In general, the
ranks $n_{1},n_{2}$ of $O_{1}$ and $O_{2}$ are different, so that $\mathbf{G}$
has $n_{1}+n_{2}$ indexes. For the same parity of $O_{1},O_{2}$ there will be
no $\varepsilon_{\mu\nu\lambda\rho...}$ tensors.

As it was discussed at length in \cite{Migdal11} the Pad\'{e} approximant. is
nothing but a continued fraction summing up Taylor expansion near
$t=-\Lambda^{2}$ in deep Euclidean region. In case there are tensor indexes,
we can still treat $\mathbf{G}$ as analytic function of $t$ with fixed unit
vector $\widehat{k}$. There are some conspiracy relations for these scalar
functions in front of products of $\widehat{k}$ which are needed to remove
kinematical singularities at $p=0$ of the unit vector. In case of even number
$2m$ of $\widehat{k}$ factors (same parity of $O_{1},O_{2}$ ) the scalar
function in front must vanish as $t^{m}$ otherwise it should have extra factor
of $\sqrt{t}$. In order to avoid the kinematical singularities we must single
out the factor $t^{n}$ and build Matrix Pad\'{e} approximant. for
$t^{-n}\mathbf{G}$ . It will have these kinematical poles at $t=0$ which will
now cancel by $t^{n}$.

The Pad\'{e} equation for $\mathbf{Q}$ become matrix equation in this space of
invariant tensors.
\[
0=\oint_{R}ds\frac{s^{-n}\mathbf{G}(s)\cdot\mathbf{Q}(s)}{\left(  \Lambda
^{2}+s\right)  ^{L+1}}.
\]

where $L=M+1,...,M+N$ and the contour $R$ encloses the positive real axis
clockwise (i.e. goes backwards from $+\infty$ to $0$ along the lower side of
the cut, then forward from $0$ to $+\infty$ along the upper side of the cut.

The Pad\'{e} equations determine $\mathbf{P}(t),\mathbf{Q}(t)$ up to arbitrary
right multiplication by a tensor independent of $t$. This tensor can depend on
unit vector $\widehat{k}$. This gauge invariance does not affect the matrix
product $\mathbf{P}(t)\cdot\mathbf{Q}^{-1}(t)$, which is our matrix Pad\'{e}
approximant. The problem of solution of Pad\'{e} equations involve at some
stage the problem of infinite matrix inversion for $\mathbf{Q}$ in this
Hilbert space.

In the leading conformal approximation, however, this problem of infinite
matrix inversion dramatically simplifies, because the basis is known where the
$\mathbf{G}$ matrix is block diagonal. This is the basis of irreducible
conformal tensors. Corresponding 2-point functions $\mathbf{G}(t)$ are
diagonal in all conformal quantum numbers, i.e. $O(d)$ quantum numbers plus
scaling dimensions. In particular, in case of two symmetric traceless tensors
of the same dimension $\Delta$ and the same rank $n$ in coordinate space
(different ranks or different dimensions do not correlate)%
\begin{equation}
\widetilde{\mathbf{G}}(x)=\left(  \prod_{i=1}^{n}\left(  x^{2}\delta_{\mu
_{i}\nu_{i}}-2x_{\mu_{i}}x_{\nu_{i}}\right)  \right)  _{sym}\left(
x^{2}\right)  ^{-n-\Delta};
\end{equation}

where $sym$ denotes symmetrization and subtraction of traces for $\{\mu_{i}\}$
indexes as well as $\{\nu_{i}\}$ indexes. In momentum space (with
$\gamma=\Delta-d/2$)%
\begin{equation}
\mathbf{G}(k)=\sigma\left(  n+\gamma\right)  \left(  \prod_{i=1}^{n}\left(
-\partial^{2}\delta_{\mu_{i}\nu_{i}}+2\partial_{\mu_{i}}\partial_{\nu_{i}%
}\right)  \right)  _{sym}\left(  k^{2}\right)  ^{n+\gamma}%
;\label{MomentumSpaceSpin}%
\end{equation}
\qquad After all differentiations and symmetrizations we get
\begin{equation}
\mathbf{G}(k)=(k^{2})^{\gamma}\sum_{l=0}^{n}g_{l}\mathbf{T}_{l};
\end{equation}

with complete set of invariant tensors $\mathbf{T}_{l}$ and calculable
coefficients $g_{l}$ in front (see Appendix2). Explicit form of these tensors
\begin{align}
\mathbf{T}_{0}  &  =\left(  \prod_{j=1}^{n}\delta_{\mu_{j}\nu_{j}}\right)
_{sym},\\
\mathbf{T}_{l}  &  =\left(  \prod_{i=1}^{l}\widehat{k}_{\mu_{i}}%
\widehat{k}_{\nu_{i}}\prod_{j=l+1}^{n}\delta_{\mu_{j}\nu_{j}}\right)
_{sym},\\
\widehat{k}  &  =\frac{k}{|k|}.
\end{align}

Let us now multiply $\mathbf{G}(k)$ by $\mathbf{Q}(k)$. First, we expand
$\mathbf{Q}(k)$ in the same set
\begin{equation}
\mathbf{Q}(k)=\sum_{m=0}^{n}\mathbf{T}_{m}Q^{(m)}(t);t=-k^{2}%
\end{equation}
By construction these invariant tensors $\mathbf{T}_{m}$ satisfy algebraic
relation, with some Clebsch coefficients
\begin{equation}
\mathbf{T}_{l}\cdot\mathbf{T}_{m}=\sum_{j=0}^{n}C_{lm}^{j}\mathbf{T}_{j}%
\end{equation}
Note that $\operatorname*{Tr}\mathbf{T}_{i}\cdot\mathbf{T}_{j}\neq\delta_{ij}$
so that matrix inversion is needed to compute Clebsch coefficients from this relation.%

\begin{align*}
T_{i}  &  =\operatorname*{Tr}\mathbf{T}_{i},\\
T_{ij}  &  =\operatorname*{Tr}\mathbf{T}_{i}\cdot\mathbf{T}_{j},\\
T_{ilm}  &  =\operatorname*{Tr}\mathbf{T}_{i}\cdot\mathbf{T}_{l}%
\cdot\mathbf{T}_{m}.
\end{align*}%
\[
C_{lm}^{j}=T_{ji}^{-1}T_{ilm}.
\]
In particular%
\begin{align*}
T_{il0}  &  =T_{il},\\
C_{l0}^{j}  &  =\delta_{jl}%
\end{align*}
These matrices $T_{ij},T_{ilm}$ are independent of $\widehat{k}$ in virtue of
$O(d)$ invariance. We shall also use these Clebsch coefficients with lower
indexes%
\begin{equation}
C_{ilm}=T_{ij}C_{lm}^{j}=T_{ilm}.
\end{equation}
which are easier to compute. Making use of this algebra we get the set of
equations (with $b=\gamma-n$)%
\begin{align}
0  &  =\sum_{m=0}^{n}A_{im}\oint_{R}\frac{ds}{2\pi i}\frac{(-s)^{b}Q^{(m)}%
(s)}{\left(  \Lambda^{2}+s\right)  ^{L+1}},\\
A_{im}  &  =\sum_{l=0}^{n}g_{l}T_{ilm}.
\end{align}

We already know how to solve such equations by means the Greens function from
Appendix 1. We write (with $\Lambda\equiv1$)%

\begin{align}
\frac{(-s)^{b}}{\left(  1+s\right)  ^{L+1}}  &  =\oint_{R}\frac{dt}{2\pi
i}\frac{(-t)^{b}}{\left(  1+t\right)  ^{L+1}}\mathit{K}_{b}(t,s),\\
0  &  =\oint_{R}\frac{dt}{2\pi i}\frac{(-t)^{\gamma}}{\left(  1+t\right)
^{L+1}}\oint_{R}\frac{ds}{2\pi i}\mathit{K}_{b}(t,s)\sum_{m=0}^{n}%
A_{im}Q^{(m)}(s),
\end{align}

From this equation (in the limit when $M,N\rightarrow\infty,R^{2}=2MN/\Lambda$
fixed, in units of $R$), we derive%
\begin{equation}
\oint_{R}\frac{ds}{2\pi i}\mathit{K}_{b}(t,s)\sum_{m=0}^{n}A_{im}%
Q^{(m)}(s)=X_{i}\sum_{r=0}^{\infty}\frac{(-t)^{r}}{\Gamma(r+1)\Gamma(r+b+1)}%
\end{equation}

with some yet undetermined factor $X_{i}$ . Now, expanding the entire
functions in convergent series%
\begin{equation}
Q^{(m)}(s)=\sum_{u=0}^{\infty}q_{u}^{(m)}(-s)^{u}%
\end{equation}

and comparing coefficients in front of $(-t)^{r}$ and using $\digamma_{\gamma
}(r,u)=\delta_{r,u}$ we get finite set of $n$ linear algebraic equations for
$n$ coefficients $q_{r}^{(m)}$ as functions of $r$:
\begin{equation}
\sum_{m=0}^{n}A_{im}q_{r}^{(m)}=\frac{X_{i}}{\Gamma(r+1)\Gamma(r+b+1)}%
,r=0,1,... \label{AQDEqn}%
\end{equation}
\begin{equation}
q_{r}^{(m)}=\frac{\left(  A^{-1}\ast X\right)  _{m}}{\Gamma(r+1)\Gamma
(r+\gamma-n+1)},m=0,..n
\end{equation}

The constant vector $X$ remains arbitrary here. This reflects the gauge
invariance of right multiplication of $\mathbf{Q}(k),\mathbf{P}(k)$ by
arbitrary matrix $\mathbf{W}(k)$. In our case this gauge matrix has a form%
\begin{equation}
\mathbf{W}(k)=\sum_{j=0}^{n}X^{j}\mathbf{T}_{j}.
\end{equation}

The simplest choice would be%
\begin{equation}
X_{i}=A_{i0}%
\end{equation}
so that
\begin{align}
q_{r}^{(m)}  &  =\frac{\delta_{m0}}{\Gamma(r+1)\Gamma(r+\gamma-n+1)},\\
Q^{(m)}(s)  &  =\delta_{m0}(-t)^{-\frac{\gamma-n}{2}}I_{\gamma-n}\left(
2\sqrt{-tR^{2}}\right)
\end{align}
Let us now discuss the numerator of Matrix Meromorphized 2-point function:%
\begin{align}
\mathbf{P}(k)  &  =\int_{C_{t}}\frac{ds(-s)^{-n}\mathbf{G}(s)\cdot
\mathbf{Q}(s)}{2\pi i(s-t)}\nonumber\\
&  =\int_{C_{t}}\frac{ds(-s)^{b}Q^{(0)}(s)}{2\pi i(s-t)}\sum_{m=0}^{n}%
g_{m}\mathbf{T}_{m}.\nonumber
\end{align}

Expanding in power series in $k^{2}=-t$ and integrating over $s$ we get%
\begin{align}
\mathbf{P}(k)  &  =\sum_{m=0}^{n}\mathbf{T}_{m}P^{(m)}(-k^{2}),\\
P^{(m)}(t)  &  =g_{m}\sum_{r=0}^{\infty}\frac{(-t)^{r}}{\Gamma(r+1)\Gamma
(r-\gamma+n+1)}=g_{m}(-t)^{\frac{\gamma-n}{2}}I_{-\gamma+n}\left(
2\sqrt{-tR^{2}}\right)  ,
\end{align}

\qquad

The resulting meromorphized function reads%
\begin{equation}
M[\mathbf{G},\mathbf{Q,}k]=\frac{(k^{2})^{\gamma-n}I_{-\gamma+n}\left(
2\sqrt{k^{2}R^{2}}\right)  }{I_{\gamma-n}\left(  2\sqrt{k^{2}R^{2}}\right)
}\sum_{s=0}^{n}g_{s}\left(  \prod_{i=1}^{s}k_{\mu_{i}}k_{\nu_{i}}\prod
_{j=s+1}^{n}k^{2}\delta_{\mu_{j}\nu_{j}}\right)  _{sym}%
\end{equation}
The mass spectrum $m_{n}$ defined by roots of Bessel function%
\begin{equation}
J_{\Delta-n-\frac{d}{2}}(2mR)=0.
\end{equation}

For typical tensor family in QCD with some number of covariant derivatives
$\Delta-n$ does not depend on $n$ in zeroth order of perturbation theory. This
dependence comes only from anomalous dimension as a function of running
coupling.\qquad\qquad

Above relations generalize the meromorphization equations for spin zero we
have derived in the Introduction. The formulas (\ref{QDefs}) do not rely upon
conformal symmetry. As long as we use the Mellin transform for each scalar
function in $G(k)$
\begin{equation}
(k^{2})^{-n}\mathbf{G}(k)=\sum_{m=0}^{n}\mathbf{T}_{m}\oint_{C_{F_{l}}}%
\frac{d\omega}{2\pi i}F_{m}(\omega)\left(  -t\right)  ^{\omega},
\end{equation}

we get the following representation for $\mathbf{P}(k),\mathbf{Q}(k)$%

\begin{align}
\mathbf{Q}(k)  &  =\mathbf{T}_{0}\sum_{r=0}^{\infty}q_{r}k^{2r}\\
\mathbf{P}(k)  &  =\sum_{j=0}^{n}\mathbf{T}_{j}\sum_{r=0}^{\infty}p_{r}%
^{j}k^{2r}\\
p_{\mu}^{j}  &  =\sum_{m=0}^{n}\oint_{C_{F_{m}}}\frac{d\omega}{2\pi i}%
F_{j}(\omega)q_{\mu-\omega}^{(m)}.
\end{align}

Equation for $\mathbf{Q}(k)$ will change in case of running coupling constant.
In the same way as we did it for spin zero case we can use the perturbation
expansion in dimensional regularization in matrix form and relate terms of
expansion to the corrections in general Pad\'{e} equations for $\mathbf{Q}(k)$.

\qquad\qquad

\section{Acknowledgement}

\bigskip

I am grateful to Sasha Polyakov and Ed Witten for useful discussions which
helped me understand relation between old and new theories. I would also like
to thank my son Arthur for his help with drawings and slide shows. Recently, I
discussed this paper with Marco Bochicchio, which helped me clarify the
confinement picture. I am also grateful to Misha Shifman and other
participants of the seminar at William I Fine Theoretical Physics Institute
for their hospitality and useful discussions.

\bigskip

\section{Appendix 1. Missing Chapter of Pad\'{e} Books}

\bigskip

Let us rederive here Greens function of Pad\'{e} equations \cite{MigdalPade}
in modern notations and typeset. We express derivatives of our function
$G_{0}=(-t)^{\nu}$at $t=-\Lambda$ as Cauchy integrals and we arrive at the
following set of Pad\'{e} equations (we set the normalization point
$\Lambda=1$ to simplify formulas)%
\begin{equation}
\oint_{R}\frac{dt}{2i\sin\left(  \pi\nu\right)  }\frac{(-t)^{\nu}}{\left(
1+t\right)  ^{L+1}}Q_{0}(t)=0. \label{PadeEqns}%
\end{equation}

where $L=M+1,...,M+N$ and the contour $R$ encloses the positive real axis
clockwise (i.e. goes backwards from $+\infty$ to $0$ along the lower side of
the cut, then forward from $0$ to $+\infty$ along the upper side of the cut.%

\begin{figure}[ptb]%
\centering
\ifcase\msipdfoutput
\includegraphics[
height=3.7325in,
width=4.8706in
]%
{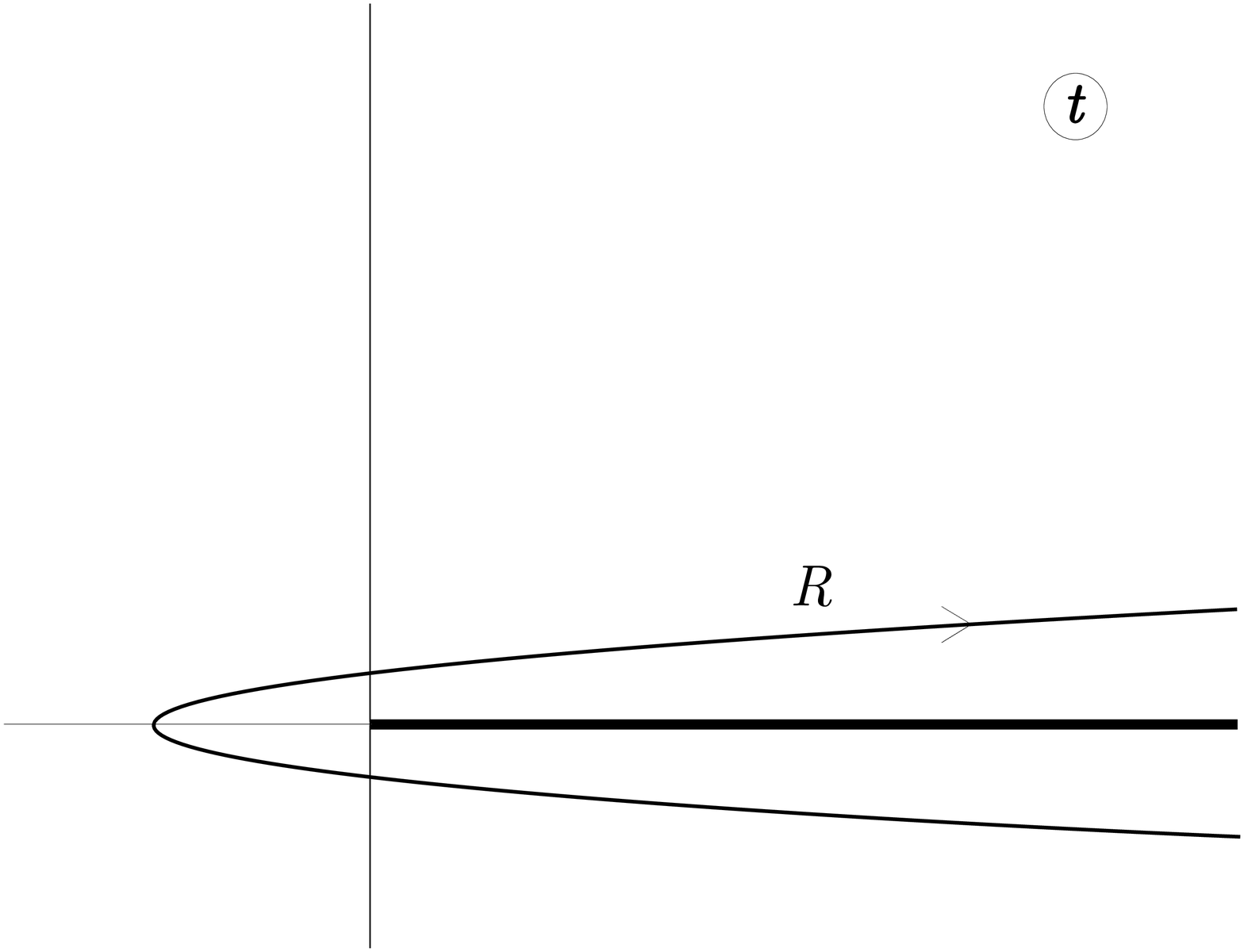}%
\else
\includegraphics[
height=3.7325in,
width=4.8706in
]%
{C:/whitepapers/Physics/Meromorphization/graphics/t-plane__3.pdf}%
\fi
\end{figure}

The discontinuity of $(-t)^{\nu}$ along this cut equals to $\sin\left(  \pi
\nu\right)  t^{\nu}$ which explains the factors in denominator. We are looking
for the Greens function $\mathit{K}(t,s)$ of Pad\'{e} equations, which must
satisfy inhomogeneous equations with proper right side:%
\begin{equation}
\oint_{R}\frac{dt}{2\pi i}\frac{(-t)^{\nu}}{\left(  1+t\right)  ^{L+1}%
}\mathit{K}_{\nu}(t,s)=\frac{\left(  -s\right)  {}^{\nu}}{\left(  1+s\right)
^{L+1}}. \label{PadeGreensFunction}%
\end{equation}

The implied extra condition is that $\mathit{K}_{\nu}(t,s)$ must be $N$-th
degree polynomial in $t$ with $s$-dependent coefficients, so that we have
linear system of integral equations for these coefficients as functions of
$s$. In the same way, $Q_{0}(t)$ is $N$-th degree polynomial in $t$ with
constant coefficients.

The general Pad\'{e} equation, with arbitrary $G(s)=(-s)^{\nu}(1+g(s))$ reads
\begin{equation}
\oint_{R}ds\frac{\left(  -s\right)  ^{\nu}}{\left(  1+s\right)  ^{L+1}%
}(1+g(s))Q(s)=0.
\end{equation}

where $Q(s)$ is $N$-th degree polynomial. Replacing $\frac{(-s)^{\nu}}{\left(
1+s\right)  ^{L+1}}$ by the left side of \ref{PadeGreensFunction}, this
equation can be expressed as an integral equation using $K,Q_{0}$:%
\begin{equation}
\oint_{R}\frac{ds}{2\pi i}\mathit{K}_{\nu}(t,s)\left(  1+g(s)\right)
Q(s)=Q_{0}(t). \label{GreensFunction}%
\end{equation}

Let us check that the solution for $K$ is given by the following Mellin-Barnes
integral%
\begin{equation}
\mathit{K}_{\nu}(t,s)=\frac{(-1)^{\nu}\pi}{\sin(\pi\nu)}\oint_{C}\frac
{dz}{2\pi i}\oint_{C^{\prime}}\frac{dz^{\prime}}{2\pi i}\frac{f(z)(1+t)^{z}%
}{f(z^{\prime})(1+s)^{z^{\prime}+1}}\frac{1}{z-z^{\prime}},
\label{MellinBarnes}%
\end{equation}

where%
\begin{equation}
f(z)=\frac{\Gamma(M+N+1-z)\Gamma(-z)}{\Gamma(M+1-\nu-z)\Gamma(N+1-z)},
\label{MyF(z)}%
\end{equation}

and contour $C$ encloses the poles of $f(z)$ at $z=0,1,..N$ while contour
$C^{\prime}$ encloses the zeroes of $f(z^{\prime})$ at $z^{\prime}%
=M+1-\nu,M+2-\nu,...+\infty$.%

\begin{figure}[ptb]%
\centering
\ifcase\msipdfoutput
\includegraphics[
height=3.7325in,
width=4.798in
]%
{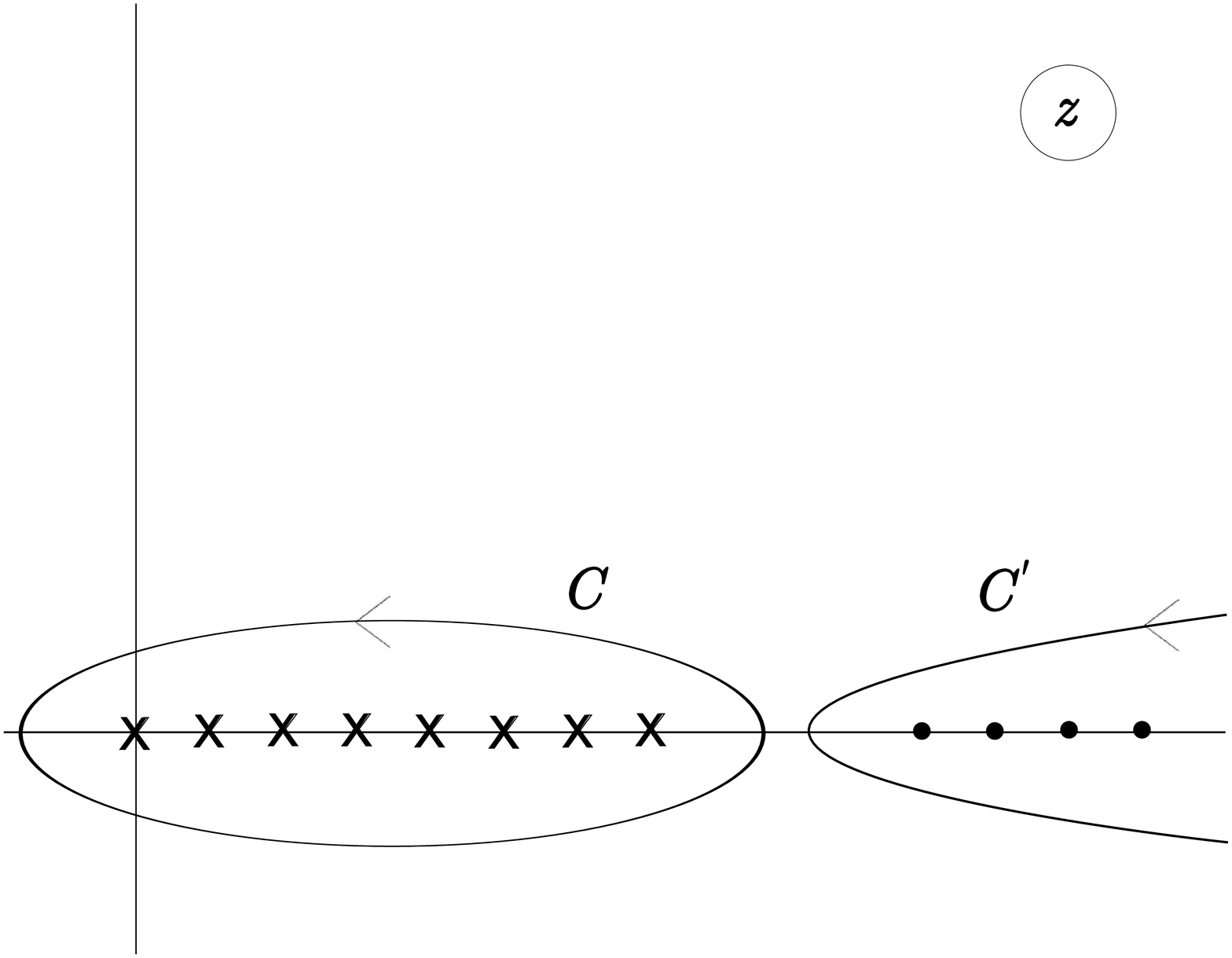}%
\else
\includegraphics[
height=3.7325in,
width=4.798in
]%
{C:/whitepapers/Physics/Meromorphization/graphics/z-plane__4.pdf}%
\fi
\end{figure}

The same function $f(z)$ determines the solution $Q_{0}$ of the homogeneous
equation%
\begin{equation}
Q_{0}(t)\propto\oint_{C}\frac{dz}{2\pi i}f(z)\left(  1+t\right)  ^{z}.
\end{equation}

Let us check this statement first. We obtain the following integral in
(\ref{PadeEqns})%
\begin{equation}
\oint_{C}\frac{dz}{2\pi i}f(z)\oint_{R}\frac{dt}{2i\sin\left(  \pi\nu\right)
}\frac{(-t)^{\nu}}{\left(  1+t\right)  ^{L+1-z}}=\oint_{C}\frac{dz}{2\pi
i}f(z)B\left(  L-\nu-z,\nu+1\right)  \label{BetaIntegral}%
\end{equation}

The function $f(z)$ was chosen in such a way that the integrand here reduces
to the rational function%
\begin{equation}
f(z)B\left(  L-\nu-z,\nu+1\right)  \propto\prod_{k=L+1}^{M+N}(k-z)\prod
_{l=M+1}^{L-1}(l-\nu-z)\prod_{n=0}^{N}(n-z)^{-1}.
\end{equation}
This function does not have poles outside integration contour $C$ and
decreases as $z^{-2}$ at infinity, therefore integral is equal to zero. On the
other hand, $f(z)$ by itself has poles at $z=0,1,...N$ which makes $Q_{0}$ a
polynomial of $N-th$ degree. So, this is a solution of Pad\'{e} equations.

When we substitute the Anzatz (\ref{MellinBarnes}) back into
(\ref{PadeGreensFunction}) we first integrate over $t$ and we get this time%
\begin{equation}
\frac{f(z)B\left(  L-\nu-z,\nu+1\right)  }{z-z^{\prime}}%
\end{equation}

which has exactly one pole at $z=z^{\prime}$ outside integration contour $C$
and decreases as $z^{-3}$ at infinity. Therefore, the $z$ integration reduces
to the residue at this pole. The factors $f(z)$ and $f(z^{\prime})$ cancel
among themselves and we get the standard integral, calculable by taking
residues at $z^{\prime}=L-\nu-n$. Summing up resulting binomial expansion we
finally get
\begin{equation}
(-1)^{\nu}\oint_{C^{\prime}}\frac{dz^{\prime}}{2\pi i}\frac{B\left(
L-\nu-z^{\prime},\nu+1\right)  }{(1+s)^{z^{\prime}+1}}=\sum_{n=0}^{\infty
}\frac{(-1)^{\nu-n}\Gamma(\nu+1)}{n!\Gamma(\nu-n+1)}\left(  1+s\right)
^{\nu-n-L-1}=\frac{(1+s)^{\nu}}{(1+s)^{L+1}}\left(  \frac{1}{1+s}-1\right)
^{\nu}=\frac{\left(  -s\right)  ^{\nu}}{(1+s)^{L+1}}, \label{RHSComputation}%
\end{equation}

which is the RHS of (\ref{PadeGreensFunction}).

Now, let both $M,N$ go to $\infty$. \ In this limit (assuming $z\sim MN$)%
\begin{equation}
f(z)\rightarrow\left(  -z\right)  ^{\nu-1}\exp\left(  -\frac{z_{0}}{z}\right)
,z_{0}=2MN.
\end{equation}
By rescaling variables (remember, there were factors of $\Lambda$ for each
$t,s$ variable, so now we switch to the units where $R=1$):
\[
z\rightarrow z_{0}z,z^{\prime}\rightarrow z_{0}z^{\prime},t\rightarrow\frac
{t}{z_{0}},s\rightarrow\frac{s}{z_{0}}%
\]
we arrive at the integrals%
\begin{align}
\mathit{K}_{\nu}(t,s)  &  \rightarrow-\frac{\pi}{\sin(\pi\nu)}\oint_{C}%
\frac{dz}{2\pi i}\oint_{C^{\prime}}\frac{dz^{\prime}}{2\pi i}\frac{\exp\left(
\frac{1}{z^{\prime}}-\frac{1}{z}+tz-sz^{\prime}\right)  (-z)^{\nu-1}}{\left(
z-z^{\prime}\right)  (z^{\prime})^{\nu-1}},\\
Q_{0}(t)  &  \rightarrow\oint_{C}\frac{dz}{2\pi i}(-z)^{\nu-1}\exp\left(
-\frac{1}{z}+tz\right)  =\sum_{n=0}^{\infty}\frac{(-t)^{n}}{n!\Gamma(n+\nu+1)}%
\end{align}

In this paper we use the expansion coefficient\ $\mathit{K}_{\nu,m}(s)$ of
$\mathit{K}_{\nu}(t,s)$ in front of $(-t)^{m}$, integrated over $s$ with
weight $(-s)^{\omega}$, corresponding to Mellin transform%
\begin{equation}
\oint_{R}\frac{ds}{2\pi i}\mathit{K}_{\nu,m}(s)(-s)^{\omega}=-\frac{\pi}%
{\sin(\pi\nu)m!\Gamma(-\omega)}\oint_{C}\frac{dz}{2\pi i}\oint_{C^{\prime}%
}\frac{dz^{\prime}}{2\pi i}\frac{\exp\left(  \frac{1}{z^{\prime}}-\frac{1}%
{z}\right)  (-z)^{\nu+m-1}}{\left(  z-z^{\prime}\right)  (z^{\prime}%
)^{\nu+\omega}}%
\end{equation}

This integral simplifies by the following change of variables%
\[
z=\frac{1}{p},z^{\prime}=\frac{-1}{q}.
\]

and adding one more integration%

\begin{equation}
\int_{0}^{\infty}dx\oint_{R}\frac{dq}{2\pi i}\oint_{R}\frac{dp}{2\pi
i}(-p)^{-\nu-m}(-q)^{\nu+\omega-1}\exp\left(  -(1+x)(p+q)\right)  ,
\end{equation}

where contour $R$ encircle positive axis clockwise, as before. These integrals
reduce to Gamma functions:%
\begin{align*}
\oint_{R}\frac{dq}{2\pi i}(-q)^{\omega+\nu-1}\exp\left(  -q(1+x)\right)   &
=\frac{(1+x)^{-\omega-\nu}}{\Gamma(-\omega-\nu+1)},\\
\oint_{R}\frac{dp}{2\pi i}(-p)^{-\nu-m}\exp\left(  -p(1+x)\right)   &
=\frac{(1+x)^{m+\nu-1}}{\Gamma(m+\nu)},\\
\int_{0}^{\infty}dx(1+x)^{m-\omega-1}  &  =\frac{1}{\omega-m}.
\end{align*}

Collecting all factors we arrive at expression (\ref{DeltaSymbol})%
\begin{equation}
\oint_{R}\frac{ds}{2\pi i}\mathit{K}_{\nu,m}(s)(-s)^{\omega}=\digamma_{\nu
}(m,\omega), \label{KNorm}%
\end{equation}

Taking the limit at integer $\omega\rightarrow n\geq0,$%

\begin{equation}
\oint_{R}\frac{ds}{2\pi i}\mathit{K}_{\nu,m}(s)(-s)^{n}=\delta_{mn}.
\end{equation}

This normalization provides that at $g(s)=0$ our integral equation
(\ref{GreensFunction}) has solution $Q(s)=Q_{0}(s)$.

\bigskip

\section{Appendix 2. Conformal 2-point function in momentum space}

\bigskip

\bigskip

The momentum space 2-point function of two symmetric traceless tensors of rank
$n$ has the form (\ref{MomentumSpaceSpin}).

In order to compute these $g_{m}$ one may use generating function by
multiplying this equation by products of complex light-like vectors $\xi,\eta$.

All the trace terms disappear in virtue of $\xi^{2}=\eta^{2}=0$ and we are
left with the following polynomial
\begin{equation}
\sigma\left(  n+\gamma\right)  \left(  -(\xi\eta)\partial^{2}+2\left(
\xi\partial\right)  \left(  \eta\partial\right)  \right)  ^{n}\left(
k^{2}\right)  ^{\gamma+n}=(k^{2})^{\gamma}\sum_{m=0}^{n}g_{m}\left(
\xi\widehat{k}\right)  ^{m}\left(  \eta\widehat{k}\right)  ^{m}(\xi\eta
)^{n-m},
\end{equation}

Let us introduce the polynomial $F_{l}(x)$ as follows (we keep constant
parameter $b=\gamma+n$)%
\begin{align}
F_{l}\left(  x\right)   &  =\sigma\left(  b\right)  \left(  k^{2}\right)
^{l-b}\left(  -\partial^{2}+\frac{2\left(  \xi\partial\right)  \left(
\eta\partial\right)  }{(\xi\eta)}\right)  ^{l}\left(  k^{2}\right)  ^{b},\\
x  &  =\frac{\left(  \xi k\right)  \left(  \eta k\right)  }{(\xi\eta)k^{2}},\\
F_{n}(x)  &  =\sum_{m=0}^{n}g_{m}x^{m}.
\end{align}

Shifting $l$ by $1$ we find recurrent equation%
\begin{align}
F_{l+1}\left(  x\right)   &  =\left(  k^{2}\right)  ^{l+1-b}\left(
-\partial^{2}+\frac{2\left(  \xi\partial\right)  \left(  \eta\partial\right)
}{(\xi\eta)}\right)  \left(  k^{2}\right)  ^{b-l}F_{l}\left(  x\right)  ,\\
F_{0}(x)  &  =\sigma(b).
\end{align}

We use identities (with $\partial$ being gradient with respect to $k$ )%
\begin{align*}
\partial x  &  =\frac{1}{k^{2}}\left(  \frac{\xi\left(  \eta k\right)  }%
{(\xi\eta)}+\frac{\left(  \xi k\right)  \eta}{(\xi\eta)}-2kx\right)  ,\\
\left(  \eta\partial\right)  x  &  =\frac{(\eta k)}{k^{2}}(1-2x),\\
\left(  \xi\partial\right)  x  &  =\frac{(\xi k)}{k^{2}}(1-2x),\\
\left(  \left(  \eta\partial\right)  x\right)  \left(  \left(  \xi
\partial\right)  x\right)   &  =\frac{x(1-2x)^{2}}{k^{2}},\\
\left(  \eta\partial\right)  \left(  \xi\partial\right)  x  &  =\frac{(\xi
\eta)\left(  (1-2x)-2x(1-2x)-2x(1-2x)\right)  }{k^{2}}=\frac{(\xi
\eta)(1-2x)(1-4x)}{k^{2}},\\
\left(  \partial x\right)  ^{2}  &  =\frac{0+0+4x^{2}+2x-4x^{2}-4x^{2}}{k^{2}%
}=\frac{2x(1-2x)}{k^{2}},\\
\left(  k\partial\right)  x  &  =0,\\
\partial^{2}x  &  =\frac{2}{k^{2}}\left(  1-dx\right)  .
\end{align*}

We find%
\begin{align}
\partial^{2}\left(  k^{2}\right)  ^{b-l}F_{l}\left(  x\right)   &
=\partial\left(  2(b-l)k\left(  k^{2}\right)  ^{b-l-1}F_{l}\left(  x\right)
+\left(  k^{2}\right)  ^{b-l}\partial xF_{l}^{\prime}\left(  x\right)  \right)
\nonumber\\
&  =2(b-l)(d+2(b-l-1))\left(  k^{2}\right)  ^{b-l-1}F_{l}\left(  x\right)
+0+\nonumber\\
&  +\left(  k^{2}\right)  ^{b-l}\partial^{2}xF_{l}^{x}\left(  x\right)
+\left(  k^{2}\right)  ^{b-l}\left(  \partial x\right)  ^{2}F_{l}%
^{\prime\prime}\left(  x\right) \nonumber\\
&  =\left(  k^{2}\right)  ^{b-l-1}\left(  2(b-l)(d+2(b-l-1))F_{l}\left(
x\right)  +2\left(  1-dx\right)  F_{l}^{\prime}\left(  x\right)
+2x(1-2x)F_{l}^{\prime\prime}\left(  x\right)  \right)
\end{align}
\qquad

and%
\begin{align}
\left(  \xi\partial\right)  \left(  \eta\partial\right)  \left(  k^{2}\right)
^{b-l}F_{l}\left(  x\right)   &  =\left(  \xi\partial\right)  \left(
2(b-l)\left(  \eta k\right)  \left(  k^{2}\right)  ^{b-l-1}F_{l}\left(
x\right)  +\left(  k^{2}\right)  ^{b-l}\left(  \left(  \eta\partial\right)
x\right)  F_{l}^{\prime}\left(  x\right)  \right) \nonumber\\
&  =\left(  k^{2}\right)  ^{b-l-1}(\xi\eta)\left(
\begin{array}
[c]{c}%
\left(  2(b-l)+4(b-l)(b-l-1)x\right)  F_{l}\left(  x\right) \\
+2(b-l)x(1-2x)F_{l}^{\prime}\left(  x\right) \\
+2(b-l)x(1-2x)F_{l}^{\prime}\left(  x\right) \\
+(1-2x)(1-4x)F_{l}^{\prime}\left(  x\right) \\
+x(1-2x)^{2}F_{l}^{\prime\prime}\left(  x\right)
\end{array}
\right) \nonumber
\end{align}
$\allowbreak$

Combining these terms we get%

\begin{align}
F_{l+1}\left(  x\right)   &  =\left(  -2\left(  b-l\right)  \left(
2b+d-2l-4\right)  -8\left(  b-l\right)  \left(  l-b+1\right)  x\right)
F_{l}\left(  x\right) \\
&  +\left(  2\left(  4b+d-4l-6\right)  x-16\left(  b-l-1\right)  x^{2}\right)
F_{l}^{\prime}\left(  x\right) \\
&  +(2x-1)4x^{2}F_{l}^{\prime\prime}\left(  x\right)
\end{align}

Finally, the recurrent equation reads%

\begin{align}
F_{l+1}\left(  x\right)   &  =\left(  A_{l}^{0}+A_{l}^{1}x+\left(  B_{l}%
^{0}+B_{l}^{1}x\right)  x\frac{d}{dx}+\left(  C^{0}+C^{1}x\right)  x^{2}%
\frac{d^{2}}{dx^{2}}\right)  F_{l}\left(  x\right)  ;\\
A_{l}^{0}  &  =-2\left(  b-l\right)  \left(  2b+d-2l-4\right)  ;\\
A_{l}^{1}  &  =-8\left(  b-l\right)  \left(  l-b+1\right)  ,\\
B_{l}^{0}  &  =2\left(  4b+d-4l-6\right)  ;\\
B_{l}^{1}  &  =-16\left(  b-l-1\right)  ,\\
C^{0}  &  =-4;\\
C^{1}  &  =8;\\
F_{0}(x)  &  =\sigma(b).
\end{align}

Above differential equations provide recurrent relations which allow one to
find expansion coefficients $f_{m}^{(l)}$ one after another, for
$l=0,1,...n$.
\begin{align}
f_{m}^{(l+1)} &  =M_{m,m-1}(l)f_{m-1}^{(l)}+M_{m,m}(l)f_{m}^{(l)}%
,m=0,...l+1,\\
f_{-1}^{(l)} &  =f_{l+1}^{(l)}=0;\\
M_{m,m}(l) &  =A_{l}^{0}+mB_{l}^{0}+m(m-1)C^{0},\\
M_{m,m-1}(l) &  =A_{l}^{1}+(m-1)B_{l}^{1}+(m-1)(m-2)C^{1};
\end{align}

This is multiplication by bi-diagonal $l\times(l-1)$ matrix $\mathbf{M}(l)$
with diagonal elements $M_{m,m}(l)$ and sub-diagonal elements $M_{m,m-1}(l)$ ,
other elements being equal to zero. So, the solution of this recurrent
equation can be written as a matrix product for vector $\mathbf{g}$ with
components $g_{m}$
\begin{align}
\mathbf{g}  &  =\mathbf{f}^{(n)}=\mathbf{M}(n-1)\mathbf{M}(n-2)...\mathbf{M}%
(0)\mathbf{f}^{(0)},\\
f_{m}^{(0)}  &  =\sigma(b)\delta_{m0},\\
b  &  =\gamma+n.
\end{align}
\qquad

\end{document}